\documentclass[preprint,12pt,3p,times,twocolumn]{elsarticle}

\usepackage{amssymb}
\usepackage{amsthm}

\usepackage{lineno}

\usepackage{multirow}
\usepackage{listings}
\usepackage{hyperref}
\usepackage{algorithm2e}
\RestyleAlgo{ruled}
\usepackage[english=american]{csquotes}
\usepackage{float}
\usepackage{soul}
\usepackage{adjustbox}
\usepackage{tcolorbox}
\usepackage{caption}
\usepackage{booktabs}
\usepackage{subcaption}
\usepackage{pgfplots}
\usepackage{import}
\usepackage{amsmath}
\usepackage{diagbox}
\usepackage[numbers]{natbib}
\usepackage{pifont}
\newcommand{\xmark}{\ding{55}}
\pgfplotsset{compat = newest}
\newcommand*\GnuplotDefs{
        set samples 51;
        Binv(p,q)=exp(lgamma(p+q)-lgamma(p)-lgamma(q));
        beta(x,p,q)=p<=0||q<=0?1/0:x<0||x>1?0.0:Binv(p,q)*x**(p-1.0)*(1.0-x)**(q-1.0);
}

\definecolor{green_ok}{rgb}{0.2,0.71,0.33}

\newtheorem{definition}{\textbf{Definition}}
\newenvironment{customdef}[1]
  {\renewcommand\thedefinition{#1}\definition}
  {\enddefinition}

\newcommand{\Florian}[1]{\textcolor{blue}{{\it [Florian: #1]}}}

\newcommand{\eg}{\textit{e}.\textit{g}., }
\newcommand{\ie}{\textit{i}.\textit{e}., }

\newcommand{\rqone}{\textbf{\textit{How do DeepCrime mutations fare in a probabilistic MT setting?}}}
\newcommand{\rqtwo}{\textbf{\textit{What errors are induced by the different approximations made in the proposed framework?}}}

\journal{Information and Software Technology}

\begin{document}

\begin{frontmatter}

\title{A Probabilistic Framework for Mutation Testing in Deep Neural Networks}

\author[1]{Florian Tambon\corref{cor1}}\ead{florian-2.tambon@polymtl.ca}
\author[1]{Foutse Khomh}
\author[1]{Giuliano Antoniol}

\cortext[cor1]{Corresponding author}

%\affiliation[1]{organization={Department of Software Engineering - Polytechnique Montreal},
%            addressline={2500, chemin de Polytechnique},
%            city={Montreal},
%           postcode={H3T1J4},
%            state={Quebec},
%            country={Canada}}

% ONLY FOR PUSHING ON ARXIV
\address[a]{Department of Software Engineering - Polytechnique Montreal, 2500, chemin de Polytechnique, H3T1J4, Canada}

\begin{abstract}

\textit{Context:} Mutation Testing (MT) is an important tool in traditional Software Engineering (SE) white-box testing. It aims to artificially inject faults in a system to evaluate a test suite's capability to detect them, assuming that the test suite defects finding capability will then translate to real faults. If MT has long been used in SE, it is only recently that it started gaining the attention of the Deep Learning (DL) community,  with researchers adapting it to improve the testability of DL models and improve the trustworthiness of DL systems.

\textit{Objective:} If several techniques have been proposed for MT, most of them neglected the stochasticity inherent to DL resulting from the training phase. Even the latest MT approaches in DL, which propose to tackle MT through a statistical approach, might give inconsistent results. Indeed, as their statistic is based on a fixed set of sampled training instances, it can lead to different results across instances set when results should be consistent for any instance.

\textit{Methods:} In this work, we propose a Probabilistic Mutation Testing (PMT) approach that alleviates the inconsistency problem and allows for a more consistent decision on whether a mutant is killed or not.

\textit{Results:} We show that PMT effectively allows a more consistent and informed decision on mutations through evaluation using three models and eight mutation operators used in previously proposed MT methods. We also analyze the trade-off between the approximation error and the cost of our method, showing that relatively small error can be achieved for a manageable cost.

\textit{Conclusion:} Our results showed the limitation of current MT practices in DNN and the need to rethink them. We believe PMT is the first step in that direction which effectively removes the lack of consistency across test executions of previous methods caused by the stochasticity of DNN training.

\end{abstract}

%%Graphical abstract
%\begin{graphicalabstract}
%\includegraphics{grabs}
%\end{graphicalabstract}

%%Research highlights
%\begin{highlights}
%\item Research highlight 1
%\item Research highlight 2
%\end{highlights}

\begin{keyword}
%% keywords here, in the form: keyword \sep keyword

%% PACS codes here, in the form: \PACS code \sep code

%% MSC codes here, in the form: \MSC code \sep code
%% or \MSC[2008] code \sep code (2000 is the default)
Deep learning \sep Mutation Testing \sep Bayesian Probability
\end{keyword}

\end{frontmatter}

% COMMENT IF PUSHING ON ARXIV
%\linenumbers

%% main text
\section{Introduction}\label{sec:intro}

Artificial Intelligence  (AI) and Machine Learning (ML) are gaining traction with countless applications, Deep Neural Networks (DNN) being one of the most prominent components. DNN provides unprecedented capability, tackling complex classification and regression tasks, especially in computer vision. Nonetheless, they also pose new verification and validation challenges \cite{Marijan19}.  DNN behavior is dictated by its internal logic, a logic not coded by a human, but  \enquote{learned} from data.

In traditional software development, testing is an essential set of activities aiming to identify defects and verify/validate that a system meets specific requirements \cite{Jamil16}. However, despite the effort to adapt traditional software testing techniques \cite{Shahid11} to the new DNN paradigm \cite{Pei19}, to the best of the authors' knowledge, there is no convincing proof of real effectiveness in improving DNN dependability. Indeed, the stochastic nature of DNN challenges traditional software testing approaches.

%Techniques such as code coverage \cite{Miller63}, fault-injection \cite{Carreira99}, static analysis \cite{Wichmann95} or any methods ensuring some form of artifact control \cite{Spanoudakis05} or traceability \cite{Cleland14} are some examples of established methods in software testing.

%Testing is an important aspect of software engineering which aims to validate that a system meets specific requirements \cite{Jamil16}. Techniques such as code coverage \cite{Miller63}, fault-injection \cite{Carreira99}, static analysis \cite{Wichmann95} or any methods ensuring some form of artifact control \cite{Spanoudakis05} or traceability \cite{Cleland14} are some examples of established methods in software testing.

%However, the rise of Deep Neural Networks (DNN) has brought challenges to the way testing used to be applied \cite{Marijan19}. As DNN internal logic is not coded by a human, as it would be the case for a traditional program in SE, but is \enquote{learned} from data for example using back-propagation, traditional methods for testing need to be adapted to fit this new paradigm.

Mutation Testing (MT) \cite{DeMillo78} is a proven technique in Software Engineering (SE); it is the de facto standard to compare different testing criteria \cite{Briand06,Papadakis2018} or to evaluate the quality of a test set \cite{Papadakis2018}. MT's basic assumption is that if a program $P$ and its mutated version $M$, obtained by introducing a small artificial change to $P$, differ on an input $x$ (\ie $P(x) \neq M(x)$) then the mutant $M$ is killed, that is a defect was detected. This allows establishing the performance of a test suite, assuming it will then transfer to real faults as well as comparing different testing criteria.

MT appealing idea has been initially applied to DNN  to assess test data effectiveness and detect mutated DNN by works such as \cite{Ma18, Hu19, Xie15}. However, Jahangirova et al. \cite{Jahangirova20}, argued that the DNN stochastic nature imposes an MT reformulation. Given a DNN trained instance $N_i$ and its mutant $M_j$, it is hard to assess whether for a given input $x$, $N_i(x) \neq M_j(x)$ is caused by the input discovering the mutant or simply a result of the stochastic training process. It is well known that for a  given model, architecture, hyper-parameters, train, and test sets, two trained instances $N_1$ and $N_2$ will exhibit different results on a set of inputs.
%In other words: $\exists x, N_1(x) \neq N_2(x)$.
To overcome this limitation  Jahangirova et al. \cite{Jahangirova20} proposed to adopt a statistical testing procedure where $n$ trained instances of a DNN $ \{N_1, ..., N_n\}$ are compared against $n$ trained instances of a mutated DNN $\{M_1, ..., M_n\}$ over their accuracy on the test set using a statistical test. The decision is no longer based on a single instance but rather on the distribution of instances. In summary, instead of a point-wise decision, the new criteria are based on the distribution of tests results and quantify the effectiveness of the test set to kill mutants on \textit{any} instances of the DNN. A tool and a replication package have been made available, including a set of real-faults-based mutation operators, DeepCrime \cite{Humbatova21}.

%based on $n$ training instances of both $N = \{N_1, ..., N_2\}$ and $M = \{M_1, ..., M_n\}$, which should mitigate the randomness. This led to DeepCrime \cite{Humbatova21}, a tool for MT of DNN using mutation operators that are more in line with realistic errors that developers would make, contrary to the previous mutation tools such as DeepMutation++ \cite{Hu19}.

We concur that MT needs to be adapted and the decision should not be based on a single instance, however, we also argue that the approach such as in DeepCrime \cite{Humbatova21} is limited. In fact, when comparing  $n$ healthy (\ie{} non-mutated) DNN instances and $n$ mutated DNN instances, the decision (whether or not the test set kill the mutation) depends on the given set of DNN instances (both the $n$ healthy and $n$ mutated). This is to say, if we keep everything constant but we change the instance sets, the decision may change when it should not. Worse, it can even be the case that, by chance, such an approach may declare a DNN mutated when comparing against itself, through the choice of healthy instances, which raises an interesting problem as we are not able to recognize the entity identity. In a nutshell, we argue that current existing MT frameworks in the context of DNN, due to the inherent randomness of the paradigm, resembles a sort of flaky test \cite{Zheng21}; meaning that different mutation test results may be returned upon a new test run for the same test set.

In this paper we propose a  Probabilistic MT (PMT) framework, adapting MT to DNN in the context of Bayesian estimation. PMT exploits Bayesian estimation to define a probabilistic decision criterion to identify mutated models.

\begin{comment}
In a nutshell, we maintain that given a set of model $ N=\{N_1, ..., N_{n^*}\}$, a set of mutated model $M=\{M_1, ..., M_{n^*}\}$, such as $n^* > n$,
and a test set $T$, one should sample two subsets of size $n$, one from $N$ and one from $M$, and compare them. By repeating the experience several times, one can then estimate the distribution of the probability of the test returning \enquote{mutant} for any two models' instances. To evaluate our  PMT framework, we formulate the following  Research Questions  (RQ):
\begin{itemize}
  \item [\textbf{RQ1:}] \rqone
  \item [\textbf{RQ2:}] \rqtwo
\end{itemize}
\end{comment}

We evaluate our PMT framework using three models/datasets and eight mutation operators and show how our proposed approach can alleviate the flakiness issue. The goal of this evaluation is to provide evidence that previous MT methods iterations have some consistency issues across multiple test executions and show how it can be tackled using PMT. We also investigate the trade-off between the approximation error and the number of training instances (the cost) required for computation of PMT, by repeating experiments multiple times with different sampled populations of different sizes.

This paper makes the following contributions:
\begin{itemize}
    \item A new probabilistic framework for MT which accounts for the stochasticity of DNN, with a replication package \cite{rep_pack} that can be easily adapted to any new models/mutations/datasets.
    \item An analysis of the mutation operators with PMT and a comparison to simple MT.
    \item An empirical analysis of the trade-off between the number of instances required for the testing and the approximation error.
\end{itemize}

\textbf{The rest of the paper is organized as follows:} We first present a concrete motivating example to illustrate the flakiness issue occurring in current DNN MT frameworks in Section \ref{sec:motiv_ex}. We then define the problem tackled by our approach in Section \ref{sec:methodology-prob}. In Section \ref{sec:methodology}, we introduce our probabilistic framework, a potential decision function leveraging the probabilistic approach, as well as a way of estimating the error caused by the limited number of samples. In Section \ref{sec:results}, we elaborate upon our experiments and results. Section \ref{sec:threats} discusses threats to the validity of our work. Related works are described in Section \ref{sec:related}. Finally, Section \ref{sec:conclusion} concludes the paper and discusses some future works.

\section{Motivating Example}\label{sec:motiv_ex}

\subsection{Experiment}

To understand the need for PMT let us replicate the MT process. We followed the approach proposed in DeepCrime on one model and mutation operator, as they provided a comprehensive replication package and their tool constitutes one of the latest iterations of MT to date. Bear in mind, that we exactly replicated DeepCrime process as provided in their replication package. We chose the model using MNIST \cite{LeCun98}, with the same DeepCrime architecture and hyper-parameters, as well as the \textit{delete\_training\_data} mutation operator (which removes a percentage of the data proportionally for each class). The choice of model/operator does not matter, as similar behavior occurs for any model/operator we tested on (see Section \ref{sec:results}).

First, we built and trained multiple sets of 200 model instances. A first set is the set of healthy instances (\ie non-mutated); we then produce five different sets of instances applying the mutation \textit{delete\_training\_data} with magnitudes ranging from $3.12$ to $30.93$, magnitudes being used by DeepCrime. At the end of the process, we obtain 1200 model instances. Finally, let us perform six experiments applying exactly DeepCrime statistical test (see Equation \ref{dc_test}) to assess whether a mutant is killed or not, using the same test set in all cases.

%In their paper, they used $n = 20$ training instances, yet we will also evaluate an increasing number of $n$ to assess if the number of instances in the test does influence the results. We trained $200$ instances of the original (\enquote{healthy}) model and $200$ for both mutation parameters (hence, we have $600$ training instances in total). To demonstrate the flakiness of test cases generated using the DeepCrime framework, we designed the following two experiments:

\begin{figure}
\centering
\includegraphics[width=\columnwidth]{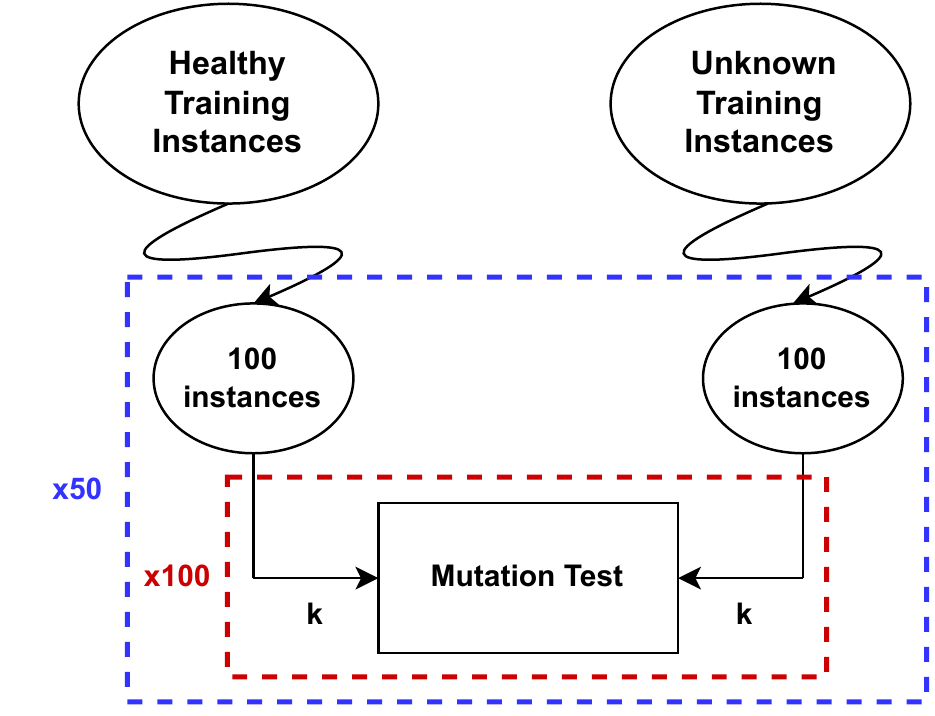}
\caption{Replication of DeepCrime's mutation test with different instances. \enquote{Unknown} means either \enquote{Healthy} or \enquote{Mutant}.} %dashed lines are approximate bagged posteriors with }
\label{fig:exp_intro_meth}
\end{figure}

A description of the example can be found in Figure \ref{fig:exp_intro_meth}. In the first experiment, we divided \enquote{healthy} instances into two disjoint sets of $100$. We pretended one of the two \enquote{healthy} set contains \enquote{unknown} instances.
We then randomly sampled $k$ instances from the $100$ \enquote{healthy} subset and $k$ out of the \enquote{unknown} set and compared them, where $k = 20$ similarly to DeepCrime's method. We repeated the sampling 100 times. We then averaged the number of times each \enquote{unknown} sample was declared \enquote{mutant} according to the statistical test used in DeepCrime, which gives us an estimation of the probability that a given \enquote{unknown} sample will be declared \enquote{mutant}.
To avoid potential sampling effect when choosing the initial two partitions of \enquote{healthy} and\enquote{unknown}, we repeat the entire process $50$ times. For all other experiments, we did the same as above, sampling sets from the $100$ \enquote{healthy} instances but this time contrasting them with sets obtained by sampling real \enquote{mutant} instances (separating experiments for each parameter magnitude). For the rest, we applied the same procedure as in the first experiment.

\subsection{Results}

Results of the procedure can be found in Table \ref{tab:exp_intro}. Remember, if the mutation test is stable and not prone to the flakiness issue we described, we should have an averaged killing probability of 0 or 1 (within a small epsilon), that is the result of the mutation test is not reliant on the instances (both \enquote{healthy} and \enquote{mutated}) used, and it either always return that the mutation is killed or always that it is not. 

If for the mutated instances with $30.93 \%$ of train data removed it is indeed the case, it is not so for all other mutation magnitudes with the probability ranging from $0.13$ to $0.85$. Worse, the \enquote{healthy} instances, if we were to use them to see if the test would consider them as mutants, are considered as such in $6\%$ of the mutation test cases. Thus, it is clear that the current mutation test is not reliable in this form as it would imply different decisions depending on the instances one would use. If we were to put it into perspective: two users using the mutation test would end up with different results on a given mutation operator just because of the instances they trained for the test, even though the architecture of the model, the dataset, the learning process, and even the test set are the same, hence the flakiness we mentioned earlier.

\begin{table}[]
\footnotesize
\caption{Average Probability of declaring \enquote{unknown} instances mutant after applying the experiments described in Figure \ref{fig:exp_intro_meth} for both \enquote{healthy} ($\mathcal{I}$) and \enquote{mutated} (with \enquote{delete training data} mutation operator and different removal percentage) as \enquote{unknown} instances.}
    
    \centering
    \begin{tabular}{c|ccccc}
         & \multicolumn{5}{c}{Mutation parameters} \\
         \hline
         $\mathcal{I}$ & 3.12 & 9.29 & 12.38 & 18.57 & 30.93\\
         \hline
         \hline
         0.06 & 0.13 & 0.45 & 0.47 & 0.85 & 1.00 \\
         \hline
    \end{tabular}
        \label{tab:exp_intro}
\end{table}

Nonetheless, in all cases, we have some instances for which the mutation test results is that the mutation is killed. As such, if we follow MT definition, the mutation is killed. However, one can see that this answer is not satisfactory given that, for instance with the identity mutation, there is only a 6\% probability on average that it happens. Thus, we argue that the question for MT, in the context of ML, \textbf{is not as much whether the mutation is killed or not, but rather \textit{how} likely it is killed or not}. To put it into perspective, the idea is similar to traditional statistical test with the concept of \textit{p-value} and \textit{effectsize}: having a significant \textit{p-value} at a given threshold means there is some statistical difference, yet the difference can be so small that it is not \textit{practically} significant, which is why \textit{effectsize} is generally used to complement \textit{p-value}. Thus, with MT, we showed it is likely there is always some differences, yet the effect of the differences is not always the same.

\section{Problem definition} \label{sec:methodology-prob}

After illustrating concretely the issue with current MT methods through the motivating example, we will introduce a few concepts relevant to our approach.

\begin{definition}\label{def:1}
For a DNN $\mathcal{N}$, let $\mathcal{D}$ be its training dataset, $\mathcal{A}$ its architecture (layers, hyperparameters,$\ldots$) and $\mathcal{P}$ its learning process (optimizer,$\ldots$). Let $\mathcal{R}$ be the set of all possible pseudo-random number generators initialization (seeds), initial values, and random values (\eg weights initialization, order of batch data,$\ldots$). We define an \textit{instance} $f = (\mathcal{D}, \mathcal{A}, \mathcal{P}, r)$ of the DNN as the model obtained after initializing the DNN and performing the stochastic process of training it by using the initialization $r \in \mathcal{R}$ .

The set of all instances of the DNN achievable through the learning process $\mathcal{P}$ of architecture $\mathcal{A}$ over dataset $\mathcal{D}$ as:
\begin{equation*}
    \mathcal{F} = \{(\mathcal{D}, \mathcal{A}, \mathcal{P} , r) | r \in \mathcal{R}\}
\end{equation*}
\end{definition}

The essential concept here is that $r$ captures and models all the stochastic elements of the training process. For example, assuming all random values used in the training process  (\eg gradient descendant, weight initialization, and others) are derived from a pseudo-random number generator, just the initial random seed and the pseudo-random algorithm knowledge will suffice to ensure the deterministic replication of the entire process. Notice that there are infinite possible seeds and thus infinite possible concrete models (\ie instantiations) each parameterized by a seed.   
If we use the object-oriented programming paradigm as a metaphor: a DNN $\mathcal{N}$ is a class whose attributes are of type $\mathcal{D}$, $\mathcal{P}$ and $\mathcal{A}$; and any instance of it corresponds to an initialization $r$ of the attributes (the weights of the layers, the order of data batch,$\ldots$) followed by applying $\mathcal{P}$ on $\mathcal{N}$. Note that for practical purpose, despite $\mathcal{F}$ being infinite, it is represented with a finite number of bits and thus its realization (on a computer) contains a large but finite number of instances.

\begin{definition}\label{def:2}
Let $\mathcal{F}$ be the set defined in Definition \ref{def:1} for a given DNN $\mathcal{N}$. Let $\mathcal{M}$ be a mutation of the DNN $\mathcal{N}$ induced either over $\mathcal{A}$, $\mathcal{P}$ or $\mathcal{D}$. %\Foutse{may be better to also define a set of mutation operators?}\Florian{Do you think it is needed? Since we will mostly compare sound instances against instances of a precise mutation? I don't want to add too much notation, so as not to overwhelm readers}.
We note the set of all instances achievable of the mutant $\mathcal{M}$ as:
\begin{equation*}
    \mathcal{F_M} = \{(\mathcal{M}(\mathcal{D}, \mathcal{A}, \mathcal{P}), r)| r \in \mathcal{R}\}
\end{equation*}

To simplify notation and for generality, we also consider the identity mutation $\mathcal{M} = I$ that is the mutation that doesn't alter the DNN, in which case $\mathcal{F}_I = \mathcal{F}$.
\end{definition}

%The mutation considered is applied to the defined DNN and each instance will have this mutation incorporated.
For instance, $\mathcal{M}$ can be \enquote{delete 3\% of the training dataset}. In that case, $\mathcal{M}$ is induced over $\mathcal{D}$. Note that, if $\mathcal{F} \cap \mathcal{F_M} = \emptyset$, it does not mean that elements of $\mathcal{F}$ and  $\mathcal{F_M}$ disagree on all possible values and for all possible instances.  For example, it is possible that  given $(f_1, f_2) \in \mathcal{F}$ and $f_\mathcal{M} \in \mathcal{F_M}$, $\exists x$ input such as $f_1(x) = f_\mathcal{M}(x)$ and $f_2(x) \neq f_1(x)$ just as described in Section \ref{sec:intro}. Yet, $f_\mathcal{M}$ is a mutated instance. In other words, on certain inputs, two \enquote{healthy} instances may disagree while they agree with a mutated instance. We can then define MT for DNN as follows:
\begin{definition}\label{def:3}
Let $\mathcal{F}$ and $\mathcal{F_M}$ be the two sets of (non-empty and finite) instances as defined in Definition \ref{def:1} and \ref{def:2}. Let $\#$ represent the cardinality of a set. Let $\mathcal{T}$ be a test set, and $n_1, n_2$ two positive integers (non-zeros, not necessarily equal). We define $S_\mathcal{F} = \{ X \subset \mathcal{F} |\ \#X = n_1\}$ and $S_\mathcal{F_M} = \{X \subset \mathcal{F_M} |\ \#X = n_2\}$. MT for DNN is a function $Z_\mathcal{T}$ defined as:
\begin{equation*}
    Z_\mathcal{T}: S_\mathcal{F} \times S_\mathcal{F_M} \longrightarrow \{0, 1\}
\end{equation*}
\end{definition}
%\Foutse{you didn't define $Z_\mathcal{T}$!!! just its domain and co-domain!}\Florian{That is the point: so far, it was like this MT has been done. People consider sound instances vs mutant instances and returned 0 or 1 depending on the test results. But $Z_\mathcal{T}$ can be any of those tests. I just want to frame it more formally. I will add the example after.}
In other words, given two sets of models' instances, the test decides if one is a mutated version of the other. Practically speaking,
$Z_\mathcal{T}$ is a definition formulated to accommodate previously published  MT functions. For example, testing one single healthy instance $N_s$ against one (single) mutant instance $N_m$, (\ie{} traditional MT), is captured in our definition by setting $S_\mathcal{F} = \{ X \subset \mathcal{F} |\ \#X = 1\}$, \\ $S_\mathcal{F_M} = \{X \subset \mathcal{F_M} |\ \#X = 1\}$ and \\ $Z_\mathcal{T}: S_\mathcal{F} \times S_\mathcal{F_M} \longrightarrow \delta_{N_s(\mathcal{T}),N_m(\mathcal{T})}$, \\where $\delta$ is the Kronecker delta.

Similarly, DeepCrime \cite{Humbatova21} mutation test is modeled by setting $S_\mathcal{F} = \{ X \subset \mathcal{F} |\ \#X = n\}$, $S_\mathcal{F_M} = \{X \subset \mathcal{F_M} |\ \#X = n\}$ and $Z_\mathcal{T}$ :

\begin{equation}\label{dc_test}
\resizebox{0.99\hsize}{!}{$
Z_\mathcal{T} = \begin{cases}
      1 & \text{if p-value} < 0.05 ~ \text{and effectSize} \geq 0.5\\
      0 & \text{else}
      \end{cases}$}
\end{equation}

Where the \textit{p-value} is obtained by using Generalised Linear Model (GLM) \cite{Nelder72} and the \textit{effectSize} is calculated using Cohen's d \cite{Kelley12} over distributions of accuracy values obtained over test set $\mathcal{T}$.

Note that PMT results depend on the test set $\mathcal{T}$ and on applied mutation operator $\mathcal{M}$, just like in traditional software engineering, but also on the sampled and compared DNN instances  (\enquote{healthy} versus \enquote{mutated}). In practice, due to resources limitation, we don't have access to $\mathcal{F}$ and $\mathcal{F_M}$ (and neither do we have $S_\mathcal{F}$ and $S_\mathcal{F_M}$). Rather, we are working with $D_s  \subset \mathcal{F}$ and $D_m \subset \mathcal{F_M}$ %\Foutse{what is s and what is m?}\Florian{s = sound, m = mutant. Do I need to define it? It is just for notation}
representing the total number of \enquote{healthy} (respectively \enquote{mutated}) trained and available instances. Thus $Z_\mathcal{T}$ turns out, in practice, to be $Z_{\mathcal{T}_{|S, S'}}$ %\Florian{Just pointing out that, in practice, we are working on a restriction of $H$, not $H$ itself.}
where $S = \{ X \subseteq D_s |\ \#X = n'_1\}$ and $S' = \{X \subseteq D_m |\ \#X = n'_2\}$. To avoid over-complicating the notations, we will refer to $Z_{\mathcal{T}_{|S, S'}}$ and $Z_\mathcal{T}$ as $Z$, since the objective is to have an approximation of a general function over $S_\mathcal{F}$ and $S_\mathcal{F_M}$ and $\mathcal{T}$ is the same in all cases.

Ideally, this function $Z$ % no matter its actual definition,
should return $0$ if $\mathcal{M} = I$ and $1$ otherwise, or, at the very least, return consistent results across any sets of instances for the same mutation operator $\mathcal{M}$. Yet, example in Section \ref{sec:motiv_ex} showed it was not the case.

\begin{figure*}[h]
\centering
\includegraphics[width=\textwidth]{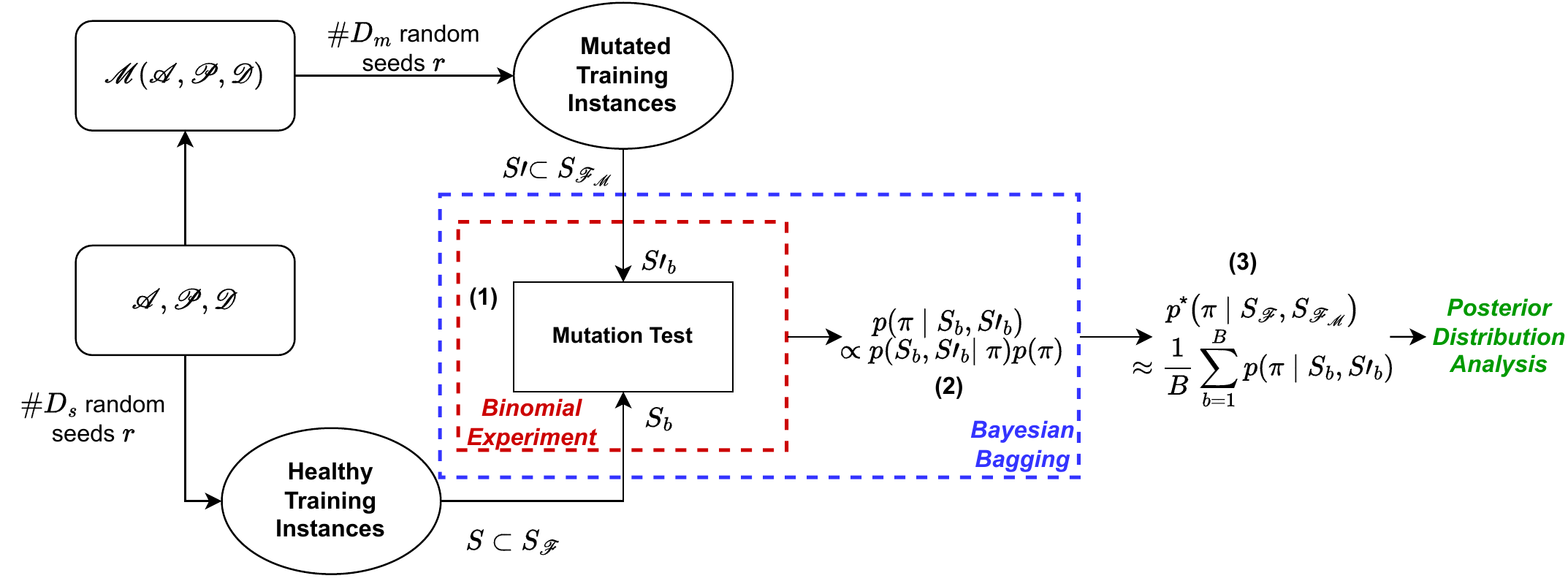}
\caption{PMT methodology overview.} %dashed lines are approximate bagged posteriors with }
\label{fig:meth}
\end{figure*}

\section{Probabilistic MT}\label{sec:methodology}

Having defined the setting we were working in, as well as showing concretely the issue of current MT, we now describe our PMT framework. An overview describing the complete process is presented in Figure \ref{fig:meth}.

\subsection{A probabilistic framework for MT}

Remember that MT compares some healthy instances of a DNN $\{N_1, ..., N_n\}$ against some mutated instances of a DNN $\{M_1, ..., M_n\}$, where $n$ is a strictly positive integer, from a pool of instances that we defined in Section \ref{sec:methodology-prob} as $S$ and $S'$. The key observation is that for any instance of $d_{s_i} \in S,\;  d_{m_i} \in S'$ computing the decision function $Z(d_{s_i}, d_{m_i})$ correspond to performing a Bernoulli trial.
%, \ie the output will be zero or one and we can assume that $H(d_{s_i}, d_{m_i})$ and $H(d_{s_j}, d_{m_j})$ are independent.

As such, instead of proposing a deterministic decision, it is possible instead to consider the probability of the outcome over $S,\;  S'$. By repeating the comparison  (\ie experiment) $N$ times, resampling at random each time, we define $Y = X_1 + ... + X_N \sim Binomial(N, \pi)$ where $X_i \sim Bernoulli(\pi)$ be the random variable representing the $i^{th}$ realization of the mutation test $Z$ (\textbf{(1)} in Fig \ref{fig:meth}).

Using Bayes rules, probability $p(\pi|S, S')$ estimation can be expressed  as a Bayesian estimation  problem for the parameter $\pi$ knowing observed data $S$ and $S'$, that is:
\begin{equation*}
p(\pi|S, S')\propto p(S, S'|\pi)p(\pi)\end{equation*}

Since $p(S, S'|\pi)$ is Binomial, we can use the Beta distribution as conjugate prior. Since we have no information on the distribution, we can use a non-informative prior. Kerman \cite{Kerman11} recommends using either the neutral prior (Beta($\frac{1}{3}$, $\frac{1}{3})$) or the uniform prior (Beta(1,1)). In our experiment, we adopted the latter choice (\ie Beta(1,1)).
%We go for the latter as it is a more established prior.
Overall,   $p(\pi|S, S')$ distribution is  $Beta(a+ k, N-k+b)$ where $a = 1, \; b = 1$ are pseudo counts and $k$ is the number of successes (see \textbf{(2)} in Fig \ref{fig:meth}).

\subsection{Bayes Bagging}\label{sec:methodology-bag}

$p(\pi|S, S')$ and thus the distribution for parameter $\pi$ are estimated over $S,\;  S'$ rather than over $S_\mathcal{F},\;  S_\mathcal{F_M}$. Remember that $S_\mathcal{F},\;  S_\mathcal{F_M}$ cannot be accessed in practice; we are limited to finite subsets. A workaround to improve estimates is to exploit \textit{Bayes Bag} \cite{Buhlmann14} which consists in applying bagging to the bayesian posterior. Huggins \cite{Huggins19} showed that Bayes Bag can result in more accurate uncertainty even with a limited number of replications ($B = 50\ \text{or}\ 100$). Furthermore,  under the assumption that a given sample is representative enough of an unknown population, we obtain,   similarly to traditional bootstrap \cite{Efron79},  an approximation of the errors of the estimates (Monte-Carlo error).

%In other words, given an original dataset $D = \{x_1, x_2, ..., x_n\}$, it is possible to generate $B$ bootstrap copies of it, \ie{} $D_i = \{x_1*, ..., x_n*\}$, $i\in \{1, 2,..., B\}$. Once copies are available, using the $B$ copies, the bagged posterior $p^*(\pi | D)$ can then be approximated as:
%\begin{equation*}
%    p^*(\pi | D) \approx \frac{1}{B}\sum_{i=1}^B p(\pi | D_i)
%\end{equation*}

In the context of PMT, using bayes bagging, we can obtain multiple bootstrapped posterior $p(\pi|S_b, S'_b)$ which can be then aggregated into $\frac{1}{B}\sum_{b=1}^B p(\pi | S_b, S'_b)$. This allows us to obtain the $p^*(\pi|S_\mathcal{F}, S_\mathcal{F_M})$ (\textbf{(3)} in Fig \ref{fig:meth}) approximation as we wanted.

%$D_b$ corresponds to $S_b, S'_b$ which are obtained through bootstrap sampling of $D_s, D_m$, that is, we first bootstrap sample from $D_s, D_m$ in order to create $D_{s_b}, D_{m_b}$ and then create $S_b, S'_b$ as done previously. We can then obtain multiple $p(\pi|S_b, S'_b)$, each $p(\pi|S_b, S'_b)$ following a beta distribution. Unfortunately,  there is no closed form for $\frac{1}{B}\sum_{b=1}^B p(\pi | S_b, S'_b)$, therefore, we  must resort to an approximation.

%Since  we know that $\pi|S_\mathcal{F}, S_\mathcal{F_M}$  follows a $Beta$ distribution, and this is true for each experiment, a different approach is to  directly compute the $Beta$ distribution $\alpha^*, \beta^*$ parameters using non-linear least square from the values of $\frac{1}{B}\sum_{b=1}^B p(\pi | S_b, S'_b)$. Overall, we estimate  $p^*(\pi|S_\mathcal{F}, S_\mathcal{F_M})$ by resolving the $\alpha^*, \beta^*$ parameters via the distribution $\pi|S_\mathcal{F}, S_\mathcal{F_M} \approx Beta(\alpha^*, \beta^*)$ (\textbf{(3)} in Fig \ref{fig:meth}).

\subsection{PMT posterior analysis} \label{sec:methodology-decision}

Once we plug the approximation of the posterior probability $p^*(\pi|S_\mathcal{F}, S_\mathcal{F_M})$ into the PMT framework, Definition \ref{def:3} is extended as:

\begin{customdef}{3.bis}\label{def:3b}
For $S_\mathcal{F}$ and $S_\mathcal{F_M}$, $\mathcal{T}$ a test set, and $Z$ a mutation function as defined in Definition \ref{def:3}. We define a probabilistic MT function $ProbZ$ as:
\begin{equation*}
    ProbZ_{\mathcal{T}, Z, N}: (S_\mathcal{F} \times S_\mathcal{F_M})^N \longrightarrow \pi|S_\mathcal{F}, S_\mathcal{F_M}
\end{equation*}
\end{customdef}

Definition \ref{def:3b} provides a means to analyze the behavior of the test set against the mutations through the analysis of the obtained posterior. Posterior can be analyzed leveraging estimates widely used in Bayesian settings, for instance:

\textbf{Point estimate:} One can derive a point estimate relaying on the posterior distribution. For instance, the Maximum A Posteriori (MAP), \ie{} $\pi^{MAP} = argmax_{\pi} p^{*}(\pi|S_\mathcal{F}, S_\mathcal{F_M})$ or the Minimum Mean Square Error (MMSE) $\hat{\pi} = \mathbb{E}(\pi|S_\mathcal{F}, S_\mathcal{F_M})$.

\textbf{Credible Interval:} A point estimate is complemented using a \textit{Credible Interval} $CI$. This is the interval within which an unobserved parameter value is present with a given probability $1 - \epsilon \%$, that is $p(\pi \in CI) = 1 - \epsilon$. Notice that,  $CI$  differs from a  \textit{confidence interval} \cite{Hespanhol19}. Multiple $CI$ exists, which can be tailored based on the point estimate used, such as the \textit{Equal-tailed} $CI$ for the median estimator, the \textit{Highest density interval} (HDI) $CI$ for the mode estimator (MAP) or the $CI$ centered around the mean. The chosen $CI$ can for instance be used in a way to measure the uncertainty of the previous probability (the wider the $CI$, the more uncertain the beliefs).

\subsection{Effect analysis}

If posterior analysis can shed some lights on the behavior of the mutations, we propose a practical criteria to establish if a mutation is \textit{likely} killed.

Remember, in the ideal case, we would like our mutation test to return either always \textit{not-mutant} or always \textit{mutant} for any instance used. The motivating example of Section \ref{sec:motiv_ex} showed it was not the case. However, the two resulting posteriors we could derive from those ideal cases can be leveraged as comparison points to calculate some form of similarity with regard to the bagged posterior obtained for a given mutation. One way to compute such probability similarity involves using the Hellinger distance \cite{cramer46} which can be defined for two beta distributions $P \sim Beta(\alpha_1, \alpha_2), ~ Q \sim Beta(\alpha_2, \beta_2)$ as:
\begin{equation*}
    H(P,Q) = \sqrt{1 - \frac{B(\frac{\alpha_1+\alpha_2}{2}, \frac{\beta_1+\beta_2}{2})}{\sqrt{B(\alpha_1, \beta_1)B(\alpha_2, \beta_2)}}}
\end{equation*}
where $B$ is the beta function.

We have that $0 \leq H(P, Q) \leq 1$, with a distance $H(P, Q)$ of $0$ implying that both $P, Q$ are the same. Thus, it's possible to calculate the distance between the bagged posterior of a given mutation and both ideal posteriors we mentioned earlier. Then, we can compute a ratio of similarity between the two distances:

\begin{equation*}\label{eq:ratio}
    \mathcal{R} = \frac{H(P,Q_H)}{H(P,Q_M)}
\end{equation*}

where $P$ is the bagged posterior, $Q_H$ the ideal posterior with all \textit{non-mutant} results and $Q_M$ the ideal posterior with all \textit{mutant} results. 

A ratio $\mathcal{R}$ of $1$ means that the bagged posterior is as similar to both ideal posterior, and so we have little information on the practical effect of the mutation is killed, which can for instance happens for a posterior centered around $0.5$ (\ie 50\% chance on average that the mutation test returns \textit{mutant} as a result for any instance). A ratio higher than $1$ implies the posterior is more similar to the ideal mutant posterior and the opposite if the ratio is lower than $1$.

In order to decide the magnitude of the effect, we elaborated the following empirical scale based of our results, inspired by existing empirical scale elaborated for \textit{effect size} criteria such as Cohen's d \cite{Sawilowsky09}, where $|d| > 1.20$ is \textit{very large}, $|d| > 0.8$ is \textit{large}, $|d| > 0.5$ is \textit{medium} and $|d| > 0.2$ is \textit{small}. In our case, we found empirically the ratio of similarity calculated with the healthy posterior to be around $0.82$ at most. Thus, since we know the healthy instances are not mutation and should not be considered as such, any ratio below $0.82$ illustrate a \textit{very strong} evidence \textit{against} the mutation being killed. From there, we can build the scale using the above-mentioned rule of thumb mirroring our ratio, with $0.82$ in our case being $1.20$ for cohen's d and $1$ in our case being their $0$. For the case above $1$, we simply take the invert of the boundaries we would get in the case below $1$. This leads to: $0.82-0.87$ (resp. $1.15-1.22$) \textit{strong}, $0.87-0.92$ (resp. $1.09-1.15$) \textit{medium}, $0.92-0.97$ (resp. $1.03-1.09$) \textit{weak}, $0.97-1.03$ \textit{negligible}. The decision is left to the user when to consider a mutation \textit{likely} killed based on the posterior or similarity ratio obtained, using some thresholds. Note that, in that configuration, we actually have three potential outcomes: the mutation is \textit{likely} killed, the mutation is \textit{likely} not killed, and no evidence points in either direction, which can happen when the thresholds are not met in either way (killed or not killed), that is we do not have enough evidence to point in either direction. In practice, this choice can default to not killing the mutation.

With what was said before, it appears that we need to redefine the traditional mutation score. %as it was defined.
The mutation score is generally defined as:
\begin{equation}
    MS = \frac{\# \text{mutations killed}}{\# \text{mutations}}
\end{equation}
that is the number of mutations killed over the total number of mutations. With our approach, we extend the mutation score to:
\begin{equation}
    MS = \frac{\# \text{mutations} ~ | ~ \mathcal{R} > \theta}{\# \text{mutations}}
\end{equation}
that is, the number of mutations for which the similarity ratio is above a certain threshold $\theta$ (\ie \textit{likely} killed) over the total number of mutations considered.

An example that leverages the complete methodology for the decision will be presented in Section \ref{sec:results-res}.

\subsection{Error estimation} \label{sec:methodology-error}

There are two types of error we aim to quantify while using PMT: the error of the bagging process (\ie if the choice of the bootstrapped data influence the bagged posterior results) and the error of the sample representativity (\ie given a certain size, does the choice of the sampled population of instances affect the bagged posterior results). The two errors will be analyzed in Section \ref{sec:results-trade-off}.

Once the bagged posterior $p^*(\pi|S_\mathcal{F}, S_\mathcal{F_M})$ is available, one can obtain error approximation. This is formulated as an estimation of a Monte-Carlo Error (MCE) \cite{Huggins19}. MCE estimation has been widely studied \cite{Koehler09}.
\\
We evaluate the MCE as follows: consider $R$ replications of our bagged posterior as Monte-Carlo simulations from which we will derive the values of Definition \ref{def:3b}. One can estimate the MCE using for instance jackknife bootstrapping \cite{Efron92}. In other words, we consider our $R$ replications of the bagged posterior $\{\pi_1|S_\mathcal{F}, S_\mathcal{F_M}, \ldots , \pi_R|S_\mathcal{F}, S_\mathcal{F_M} \}$ from which we can extract a desired estimate $e_i$ such as $X = \{e_1, \ldots,  e_R\}$.

%Let $\hat{\varphi}_R(X)$ a Monte-Carlo estimate over $X$, the jackknife estimate $J$ of the MCE is:
%\begin{equation*}
%    J(\hat{\varphi}_R(X)) = \sqrt{\frac{R - 1}{R} \sum_{q=1}^R (\hat{\varphi}_{R-1}(X_{/X_q}) - \overline{\hat{\varphi}_{R-1}(X)})^2 }
%\end{equation*}

%where $X_{/X_q}$ is the set $X$ from which the $q^{th}$ replicate was removed and $\overline{\hat{\varphi}_{R-1}(X)} = \frac{1}{R} \sum_{q=1}^R \hat{\varphi}_{R-1}(X_{/X_q})$.
We used the jackknife formula as described in \cite{Koehler09} and complemented the error estimation with a traditional confidence interval over the estimate as recommended by Koehler and Brown \cite{Koehler09}.

The estimates we will track are both the mean $\mu$ and variance $var$ of the bagged posteriors obtained. A straightforward Monte-Carlo estimator for both is simply the average over each $\mu_i$ and $var_i$ of each replicate.

\section{Evaluation of PMT}\label{sec:results}

%The \emph{goal}  of this study is to investigate whether PMT is effective in detecting DNN mutations.
%The \emph{quality focus} is the software test set effectiveness in detecting mutated DNN.
%The \emph{perspective} is that of researchers and practitioners aiming to understand the downside of current DNN MT frameworks and the advantages of PMT.
%The \emph{context} consists of three open-source datasets and seven mutation operators.
The goal of this evaluation is to investigate how much insight the PMT decision process can bring when dealing with DNN mutations and to shed some light on the limitations of current DNN MT frameworks, by comparing both frameworks and how the test is tackled in both cases. In a second time, we also analyze the trade-off between the approximation error and the cost of our method.

\begin{comment}
\begin{itemize}
\item{\bf RQ1:} \emph{\rqone} The objective of this research question is to evaluate the behavior of each mutation following the method presented in Section \ref{sec:methodology}. We also compare the selected set of mutations (see Section \ref{sec:results-data} for the selection method), and the mutation test results that would have been obtained with DeepCrime.
\item{\bf RQ2:} \emph{\rqtwo} The objective of this research question is two-fold: 1) to evaluate the monte-carlo error of the approximated bagged posterior as defined in Section \ref{sec:methodology-error}, and 2) to study the effect of selecting a limited number of training instances over the obtained bagged posterior, \ie{} the assumption that the sample is representative of the unknown population of all training instances.
\end{itemize}
\end{comment}

\subsection{Datasets and Mutations}\label{sec:results-data}

In our evaluations, we use three models/datasets, previously leveraged by DeepCrime to show how PMT alleviates the problem of flakiness mentioned earlier, and to compare it to the latest designed MT, \ie DeepCrime's definition of MT (see Equation \ref{dc_test}). More precisely, the following model/dataset combinations: MNIST \cite{LeCun98} (MN) along with a 8 layered convolutional neural network \cite{Keras1}; MovieLens dataset \cite{MovieData} to train the Movie Recommender \cite{Keras2} (MR) model; A synthesized UnityEyes (UE) dataset \cite{Wood16} along with a specific model \cite{UE}.

\begin{comment}
\begin{itemize}
    \item MNIST \cite{LeCun98} (MN) a dataset of handwritten digits widely used in deep learning, along with an 8-layered convolutional neural network \cite{Keras1}. MNIST is composed of $70,000$ images categorized into $10$ classes (0 to 9 digits). The dataset is split between $60,000$ training data and $10,000$ testing data.
    \item MovieLens dataset \cite{MovieData} is a dataset of users' movie ratings that will be used to train the Movie Recommender \cite{Keras2} (MR) model. MovieLens contains over $100,00$ ratings given by $610$ different users on over $9,700$ movies.
    \item The UnityEyes (UE) dataset along with a model \cite{UE} which learns to map images of eyes and 2D head angle to 2D eye gaze angle (yaw and pitch). This dataset was synthesized using the publicly available rendering framework of the same name \cite{Wood16}.
\end{itemize}
\end{comment}

Table \ref{tab:my_label} shows the average metric values obtained on the test set across all our \enquote{healthy} DNN trained instances; figures are in line with DeepCrime reported values. As UE and MR systems are regression-based, we considered (as in DeepCrime) that a prediction is accurate if it differs from the correct one by no more than one rating (for MR) or if the angle is no more than 5 degrees (for UE).

Regarding mutations, we chose both \textit{source-level} and \textit{model-level} mutations.

\textit{Source-level} mutations are extracted from the detailed mutation operators proposed in DeepCrime. To select which mutation to investigate in priority (and limit the number of instances to train), we used DeepCrime Killability and Triviality metrics. In their paper \cite{Humbatova21}, \textit{Killability} is defined as whether or not a mutation operator configuration is killed by the training data using the statistical test presented in Equation \ref{dc_test}. \textit{Triviality} roughly quantifies how easily a mutation operator can be killed by any test input of the test set. To push the method to its limits, for each dataset/model, we selected mutations with Killability/Triviality that is the highest/lowest possible. Finally, we also chose some mutations that are common to all models, to have %an across models
common points of comparison for all models.

Regarding \textit{Model-level} mutations, DeepCrime proposes some of those mutations but does not analyze or implement them. Therefore, we chose instead to leverage some mutation operators proposed and analyzed in MuNN \cite{Shen18} and DeepMutation \cite{Ma18}. Table \ref{tab:mut} provides an overall view of the selected mutations ($\checkmark$) for a given model/dataset, with the mutation acronym being described below:

\begin{comment}
\textit{change\_label (TCL)}, \textit{delete\_training\_data (TRD)}, \textit{change\_weights\_initialisation (WCI)}, \textit{change\_activation\_function (ACH)}, \textit{unbalance\_training\_data (TUD)}, \textit{change\_loss\_function (LCH)} and \textit{change\_optimisation\_function (OCH)}, \textit{add\_weights\_fuzzing (AWF)} and \textit{freeze\_neurons\_output (FNO)}
\end{comment}.

\begin{itemize}
    \item \textit{change\_label (TCL)}: Modify a percentage of training data labels, replacing them with the most frequent label in the dataset.
    \item \textit{delete\_training\_data (TRD)}: Remove a portion of the training dataset from each class proportionally.
    \item \textit{change\_weights\_initialisation (WCI)}: Change the way weights are initialized in all the layers of the model.
    \item \textit{change\_activation\_function (ACH)}: Change the (non-linear) activation function of a layer by another (non-linear) activation function.
    \item \textit{unbalance\_training\_data (TUD)}: Remove a portion of data belonging to the classes whose frequency of apparition is less than average.
    \item \textit{change\_loss\_function (LCH)}: Change the loss function by another loss function.
    \item \textit{change\_optimisation\_function (OCH)}: Change the optimisation function by another optimisation function.
    \item \textit{add\_weights\_fuzzing (AWF)}: Add gaussian noise of magnitude $\sigma$ to a certain percentage of weights of a layer.
    \item \textit{freeze\_neurons\_output (FNO)}: Freeze (delete) a percentage of neurons of a given layer.
\end{itemize}

DeepCrime's mutations were reused exactly as provided in the replication package. MuNN \cite{Shen18}/DeepMutation \cite{Ma18} based mutations (the two last ones) were implemented based on the description/parameters provided in the papers.
%In all cases, we strived to stay as close as possible to already published mutations and mutation parameter settings. A more detailed description is available in the  DeepCrime original paper \cite{Humbatova21}.

\begin{table*}[]
\caption{Mutations (Source / Model Level) chosen for each dataset/model. '$\checkmark$' means selected, '-' means not selected.}
    \centering
    \begin{tabular}{@{}cccccccc|cc@{}}
         \hline
         & TCL & TRD & WCI & ACH & TUD & LCH & OCH & AWF & FNO \\
         \hline
         \hline
         MN & $\checkmark$ & $\checkmark$ & $\checkmark$ & $\checkmark$ & - & - & - & $\checkmark$ & $\checkmark$ \\
         MR & $\checkmark$ & $\checkmark$ & - & - & $\checkmark$ & $\checkmark$ & - & - & - \\
         UE & $\checkmark$ & $\checkmark$ & - & - & - & $\checkmark$ & $\checkmark$ & $\checkmark$ & $\checkmark$\\
         \hline
    \end{tabular}
        \label{tab:mut}
\end{table*}

\begin{table*}[]
\caption{Systems under test. For each model, we provide the average metric value as well as standard deviation (in parenthesis).}
    
    \centering
    \begin{tabular}{@{}cccccc@{}}
         \hline
         ID & Training Data & Test Data & Epochs & Metric & Value\\
         \hline
         \hline
         MN & 60,000 & 10,000 & 12 & Accuracy & 99.15 (0.06) \\
         MR & 72,601 & 18,151 & 12 & MSE & 0.047 (0.001) \\
         UE & 103,428 & 25,857 & 50 & Angle based & 2.6$^{\circ}$ (0.2) \\
         \hline
    \end{tabular}
        \label{tab:my_label}
\end{table*}

\subsection{Instrumentation and parameters}

To carry out the experiments, we use the same requirements as documented in  DeepCrime \cite{Humbatova21}, namely,  Python (3.8), Keras (2.4.3), and Tensorflow (2.3). We also used the models/datasets, mutations operators as well as the MT procedure used in their replication package. For each mutated/healthy model, we train $200$ instances and then evaluate the accuracy of each instance on the dataset test set.

Unless specified otherwise, all experiments use the following default parameters: $N = 100$ the number of trials for each Binomial experiment and $B = 100$ the number of bootstrap repetitions. Again, such values are a compromise to ensure a trade-off between having a sufficient number of evaluations and keeping the computation within a manageable time. Moreover, we used the same number of instance as in DeepCrime ($n = 20$) for the MT with the same MT function $Z$ that they used (see Equation \ref{dc_test}).

\subsection{Experiments}

In the following, we will introduce the description of two experiments we did to evaluate our framework.

\subsubsection{First Experiment}

The first experiment aims to apply PMT to the previously listed models/mutations and to draw a comparison with MT. To do this, we leveraged  $200$ training instances per model/dataset/mutation for our method. As a point of comparison, we will apply MT on DeepCrime's instances provided in their replication package \cite{Humbatova21}.
We implemented and applied the procedure detailed in Section \ref{sec:methodology}. The procedure was needed to obtain a bagged posterior for each mutation (including the identity mutation, that is the \enquote{healthy} instances) of each dataset/model. From there, we can leverage the effect analysis method we introduced in Section \ref{sec:methodology-decision} to calculate the ratio of similarity obtained for each mutation and compare it to the results one would obtain with simple MT in order to nuance them.

\subsubsection{Second Experiment}

The second experiment aims to evaluate the error over the bagged posterior approximation and the sampled population representativity.

To estimate the bagged posterior approximation, we repeated $N_{exp} = 100$ times the calculation of the bagged posterior approximation, using the jackknife estimation as explained in Section \ref{sec:methodology-error}.

To evaluate the representativity, we considered the following. Since the trained instances are \enquote{sampled} at random when trained (\ie the random seed used in training are equally likely to be picked), the representativity of the sampled instances will depend on their number. Thus, we repeated the MCE estimation we used to estimate the bagged posterior approximation, with a different number of sample instances  (from $25$ to $190$). We repeated this process $N_{pop} = 30$ times to account for the possible effect of the choice of the samples over the obtained bagged posterior. In other words, from the
$200$ training instances, we repeated the jackknife estimation 30 times; each time with a different sampled population of the same size. This allowed us to examine the evolution of the average parameters estimate $\mu$ and $var$ across the bagged posterior as well as the average of their approximation error boundaries based on the sample size as well as the sampled instances. As described in Section \ref{sec:methodology-error}, we also compute confidence interval values as recommended in the literature \cite{Koehler09}.

\subsection{Results}\label{sec:results-res}

In this section, we only present a sample of our overall results because of space limitations. %However, , to avoid overwhelming t in the following not to overwhelm the paper with figures. 
However, we provide all the results, in our replication package \cite{rep_pack}.

\begin{figure*}
  \centering
  \caption{Posterior distribution for mutation operators of different magnitudes. Vertical dash lines symbolize the point estimate value of each posterior. Plain lines curve represent the bagged posteriors, with the colored area underneath being the \textit{CI}, while transparent lines are each posterior obtained from bootstrapped data.}
  \begin{minipage}{.45\linewidth}
    \centering
    \subcaptionbox{MR - TRD\label{fig:post-a}}
      {\includegraphics[width=\linewidth,height=0.85\textwidth]{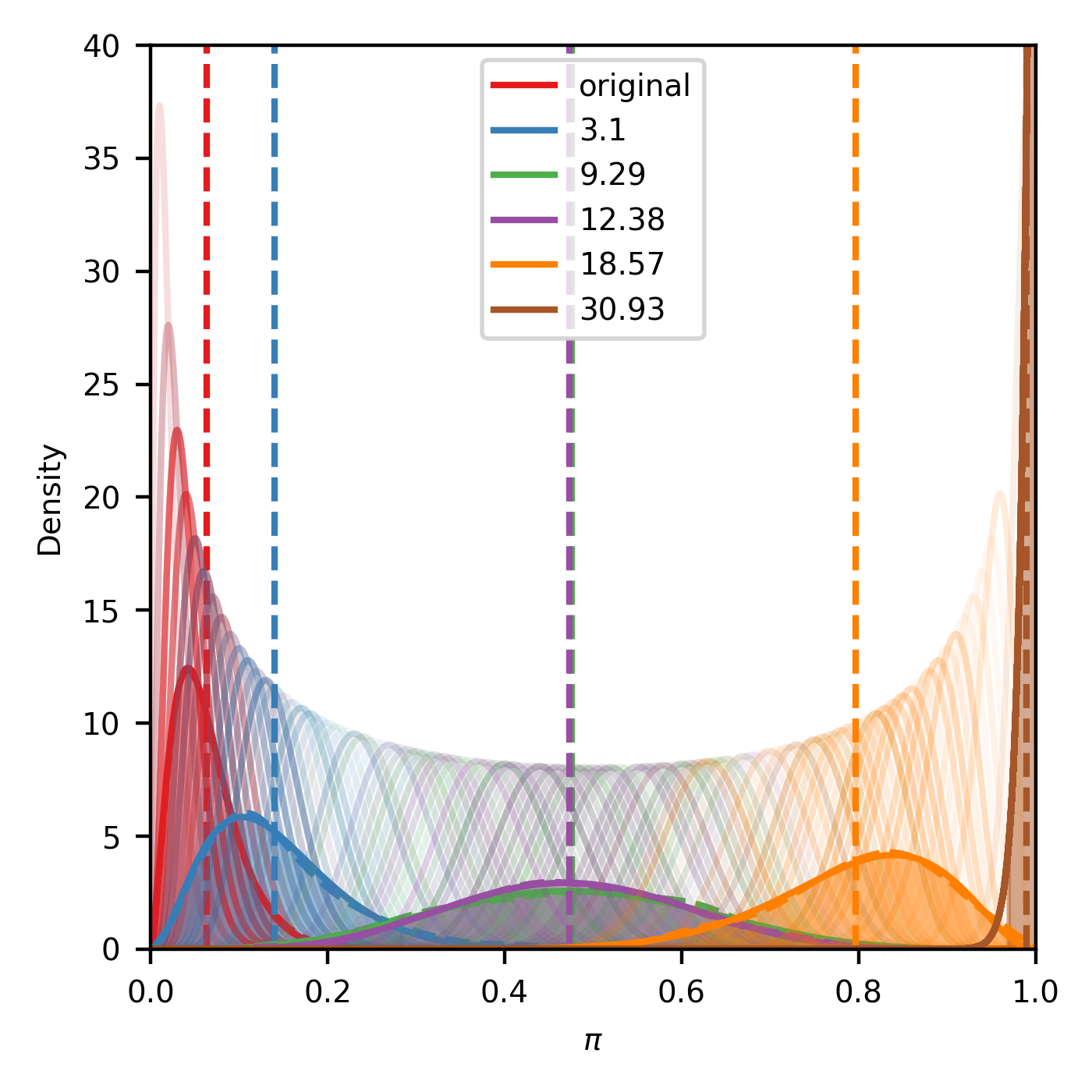}}
    \subcaptionbox{MR - TRD\label{fig:post-b}}
      {\includegraphics[width=\linewidth,height=0.85\textwidth]{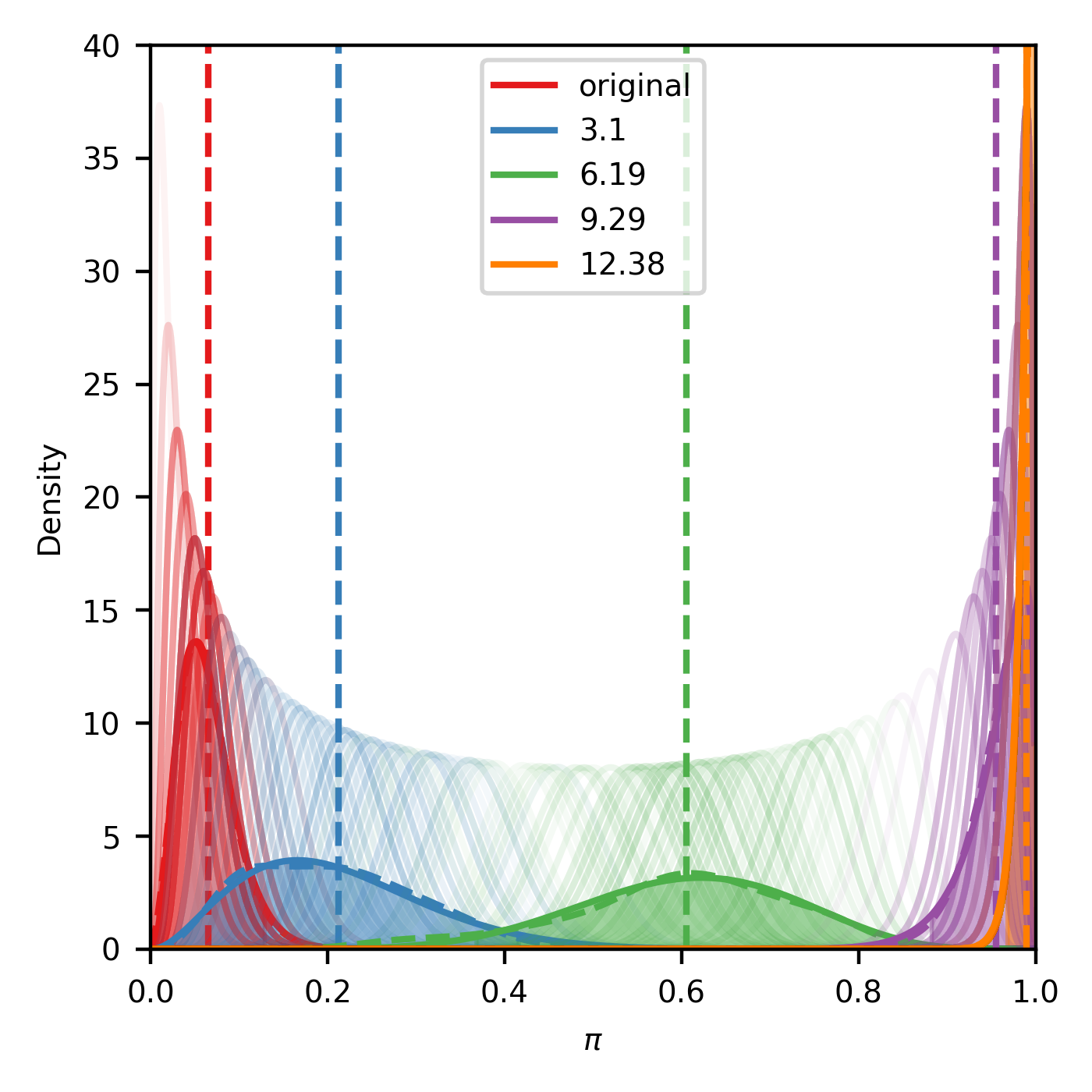}}

    \subcaptionbox{UE - TRD\label{fig:post-c}}
      {\includegraphics[width=\linewidth,height=0.85\textwidth]{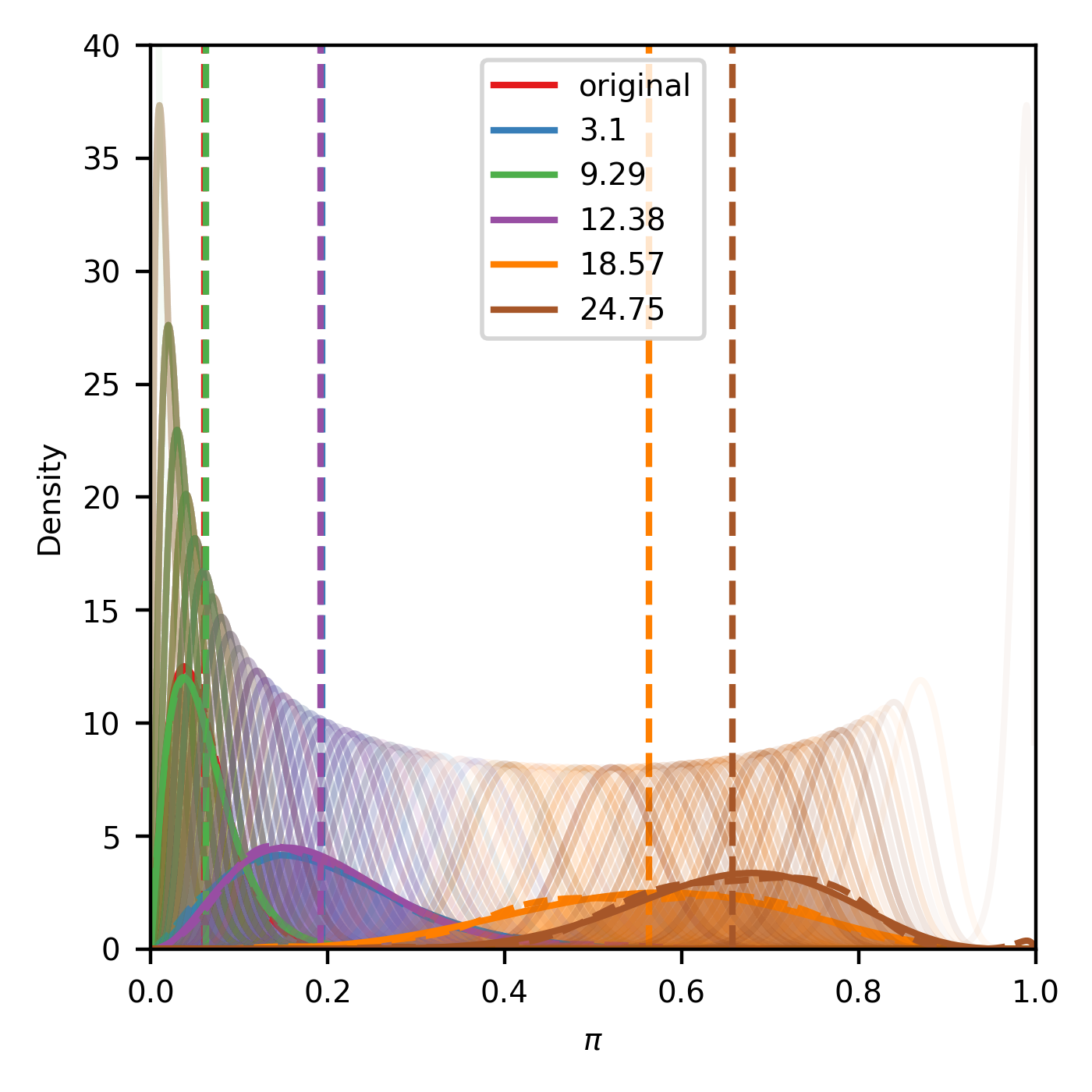}}
  \end{minipage}\quad
  \begin{minipage}{.45\linewidth}
    \centering
    \subcaptionbox{MN - ACH\label{fig:post-d}}
      {\includegraphics[width=\linewidth,height=0.85\textwidth]{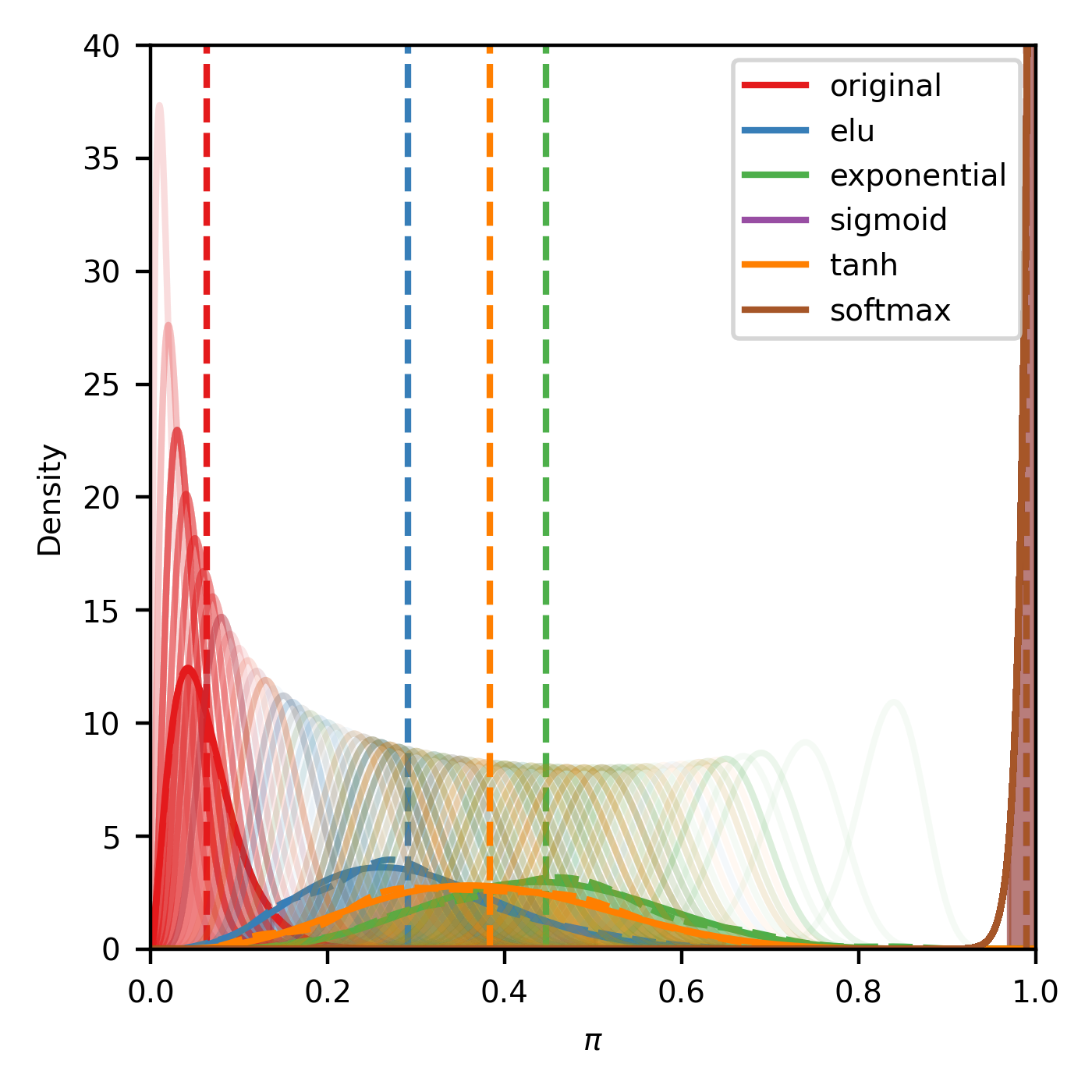}}

    \subcaptionbox{MR - TUD\label{fig:post-e}}
      {\includegraphics[width=\linewidth,height=0.85\textwidth]{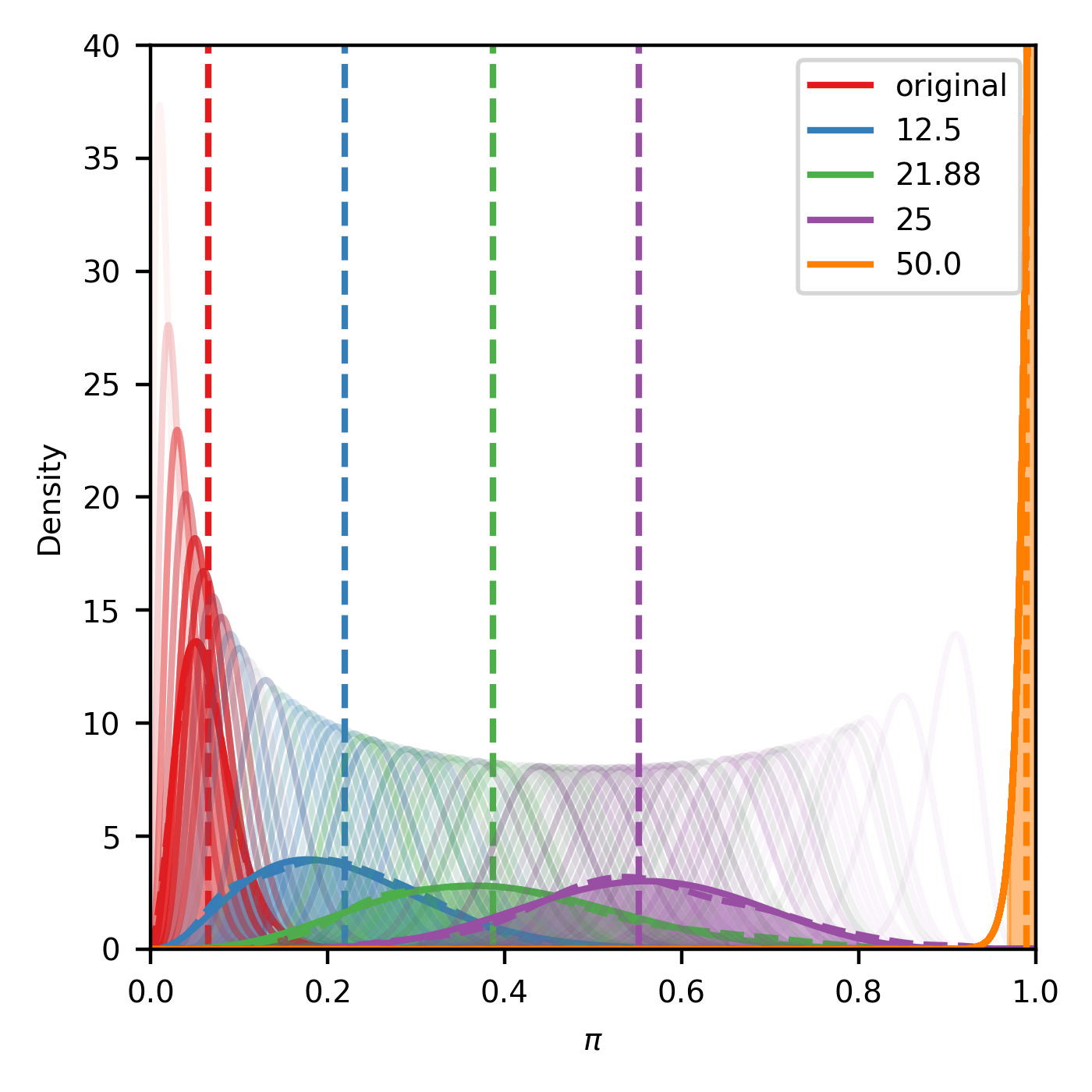}}

    \subcaptionbox{UE - TCL\label{fig:post-f}}
      {\includegraphics[width=\linewidth,height=0.85\textwidth]{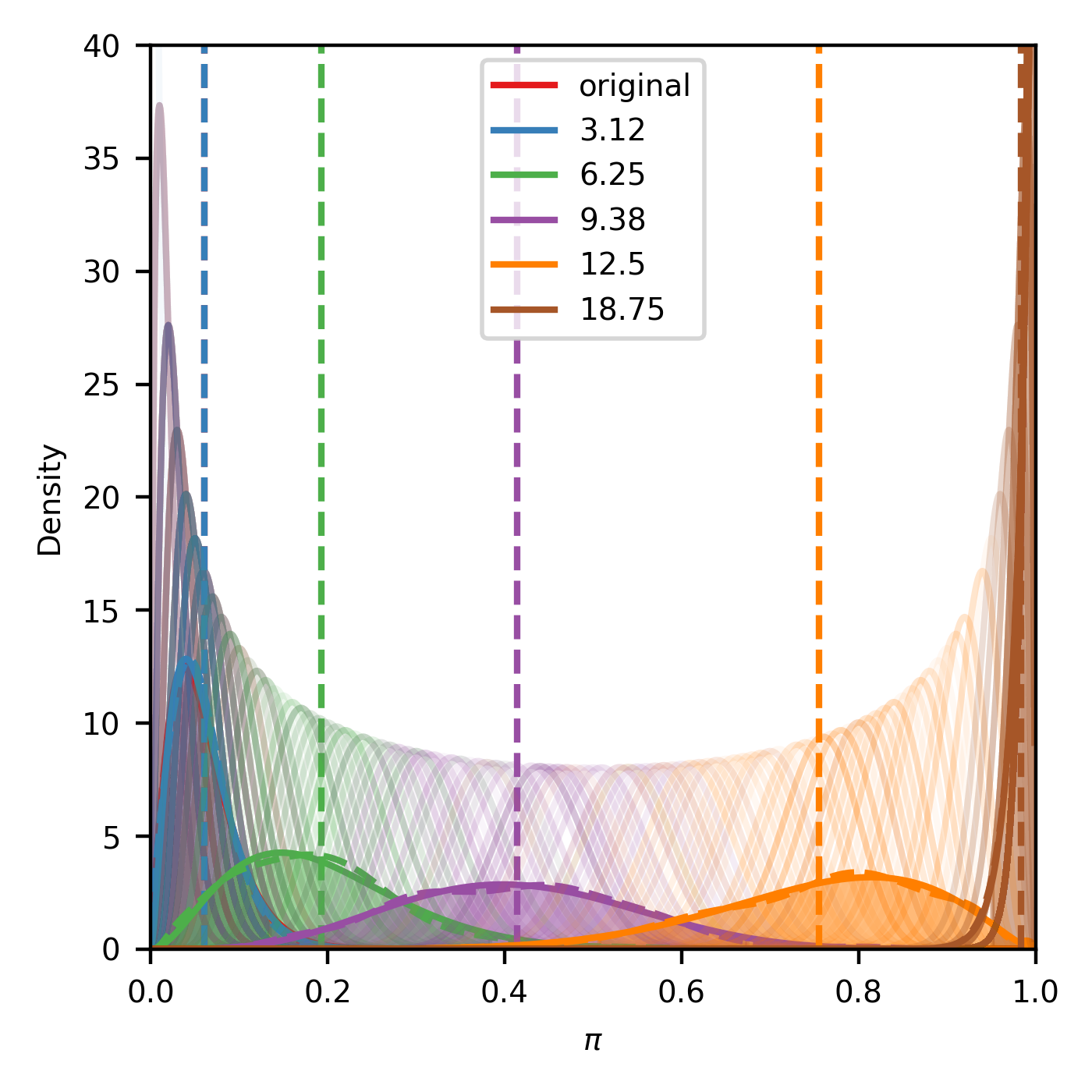}}

    \end{minipage}
    \label{fig:post}
    
\end{figure*}

\begin{table*}
\caption{Example. For MT, \enquote{\checkmark} means the mutation is killed and \enquote{\xmark} the mutation is not killed. For PMT, we give the ratio of similarity value and the effect based on the scale presented in Section \ref{sec:methodology-decision} with the following meaning: \enquote{$\circ$} \textit{negligible}, \enquote{-} \textit{weak}, \enquote{$\pm$} \textit{medium}, \enquote{+} \textit{strong} and \enquote{++} \textit{very strong}. \textcolor{red}{red} is when the effect is the likeliness of the mutation being killed ($\mathcal{R} > 1$), and \textcolor{blue}{blue} when the effect is the likeness of the mutation not being killed ($\mathcal{R} < 1$). We also present the similarity ratio when comparing healthy instances against themselves (identity mutation $\mathcal{I}$).}
    
    %\centering
    %\setlength\tabcolsep{5pt}
    \footnotesize
    \begin{tabular}{c|ccccc|ccccc}
          $\mathcal{I}$ & \multicolumn{5}{c}{MN - TRD} & \multicolumn{5}{c}{MN - ACL} \\
          0.76 & 3.1 & 9.29 & 12.38 & 18.57 & 30.93 & elu & exp & sigmoid & tanh & softmax \\
         \hline
         \hline
         MT &  \xmark & \checkmark & \xmark & \checkmark & \checkmark &  \checkmark & \checkmark & \checkmark & \checkmark & \checkmark\\
         \hline
         PMT & 0.91 (\textcolor{blue}{$\pm$}) & 1.00 ($\circ$) & 1.00 ($\circ$) & 1.05 (\textcolor{red}{-}) & $>$2 (\textcolor{red}{++}) & 0.99 ($\circ$) & 1.00 ($\circ$) & $>$2 (\textcolor{red}{++}) & 1.00 ($\circ$) & $>$2 (\textcolor{red}{++}) \\
         \hline
         \hline
         $\mathcal{I}$ & \multicolumn{5}{c}{MR - TRD} & \multicolumn{5}{c}{MR - TUD} \\
          0.81 & 3.1 & 6.19 & 9.29 & 12.38 & & 12.5 & 21.88 & 25 & 50.0 & \\
         \hline
         \hline
         MT &  \xmark & \checkmark & \checkmark & \checkmark & &  \xmark & \xmark & \xmark & \checkmark & \\
         \hline
         PMT & 0.95 (\textcolor{blue}{-}) & 1.00 ($\circ$) & 1.86 (\textcolor{red}{++}) & $>$2 (\textcolor{red}{++}) & & 0.96 (\textcolor{blue}{-}) & 1.00 ($\circ$) & 1.00 ($\circ$) & $>$2 (\textcolor{red}{++}) & \\
         \hline
         \hline
         $\mathcal{I}$ & \multicolumn{5}{c}{UE - TRD} & \multicolumn{5}{c}{UE - TCL} \\
          0.73 & 3.1 & 9.29 & 12.38 & 18.57 & 24.75 & 3.12 & 6.25 & 9.38 & 12.5 & 18.75\\
         \hline
         \hline
         MT &  \xmark & \xmark & \xmark & \xmark & \checkmark &  \xmark & \xmark & \xmark & \checkmark & \checkmark\\
         \hline
         PMT & 0.94 (\textcolor{blue}{-}) & 0.72 (\textcolor{blue}{++}) & 0.95 (\textcolor{blue}{-}) & 1.00 ($\circ$) & 1.00 ($\circ$) & 0.74 (\textcolor{blue}{++}) & 0.95 (\textcolor{blue}{-}) & 1.00 ($\circ$) & 1.05 (\textcolor{red}{-}) & $>$2 (\textcolor{red}{++}) \\
    \end{tabular}
    \label{tab:exp_1}
\end{table*}

\subsubsection{First experiment: PMT application and comparison with simple MT}\label{sec:results-exp1}

We report results for two mutation operators for each of the models, the rest of the results can be found in the replication package \cite{rep_pack}. We first report in Figure \ref{fig:post} the obtained posteriors. Each curve represents the posterior distribution for a given mutation operator magnitude following the procedure described in Section \ref{sec:methodology}. For instance, in Figure \ref{fig:post-a}, the plain orange curve represents the bagged posterior distribution of the probability of killing the \textit{delete\_training\_data} mutation with magnitude $18.57$. Its MMSE point estimate is $\hat{\pi} = 0.8$ (vertical dash line) and the credible interval width is $|CI| = 0.4$ (colored area). Transparent lines are the bootstrapped posteriors obtained from each bootstrapped data $S_b, S'_b$ (see Section \ref{sec:methodology-bag}). We report in a second time in Table \ref{tab:exp_1} a comparison between simple MT results for each of the mutation operators (\ie $1$ for the mutation is killed, $0$ if it's not) and the ratio of similarity (with the effect) as we defined in Section \ref{sec:methodology-decision} for PMT. For instance, for the mutation $MN - TRD$, the magnitude $9.29$ was considered killed by MT, yet we found a \textit{negligible} effect when using our ratio metric, \ie there is no strong argument to point out that the mutation is either \textit{likely} killed or \textit{likely} not killed. By default, the user can consider it not to be killed, in order to avoid potential false positives (\ie considering a mutation killed when it is not). Both those results will allow us to showcase the advantage of PMT over MT.

\textbf{Stability:} One thing we first showed with the motivating example and that we show again here is the lack of stability of the simple MT, \ie the flakiness we mentioned earlier. Indeed, the fact that all posterior distributions do not translate to the ideal mutant or not-mutant posterior we described (that is, MT returning always $0$ or always $1$ no matter the instances used) can have dire consequences. For instance, looking at Figure \ref{fig:post-c}, PMT shows that the posterior distribution of the mutation of magnitude $9.29$ is very similar to the healthy one. More directly, in the Table \ref{tab:exp_1}, we see that the mutation is \textit{likely} not killed with a similarity ratio of \textbf{0.72} (very strong). Nonetheless, the point estimate is non-zeros and so that means that, for some instances, there is a chance that simple MT returns \enquote{mutant} as a result, despite PMT showing strong evidence the mutation should not be considered killed. This is similar to our motivating example in Section \ref{sec:motiv_ex} where we for instance found out that 6\% of tests done on healthy instances returned mutant as a result despite no mutation being present.

\begin{tcolorbox}[colback=blue!5,colframe=blue!40!black]
\textbf{Finding 1:} PMT allows stability over test results contrary to MT. That is the decision made over a given mutation is taken while accounting for any instance possible, which prevents the flakiness issue we illustrated previously, \ie MT returning $0$ and $1$ for the same mutation depending on the instances used in the test.
\end{tcolorbox}

\textbf{Consistence:} Besides mitigating stability problems, PMT allows tackling another issue of MT: the potential lack of consistency across the tests. Indeed, as MT outcome is binary, one can not ensure that mutations that behave similarly lead to the same MT outcome, since there is no information available for the posterior of the distribution. On the contrary, using PMT, one can compare results for different mutations, whether from the same operators or from a different one. For instance, in Figure \ref{fig:post-a} both the mutation of magnitude $9.29$ and $12.38$ exhibit the same posterior and a similar ratio of similarity of \textbf{1.00} (negligible), as such, logically, a decision made over these two mutations should be the same. However, using Deepcrime's instances for MT yield opposite results, once again probably because of the training instances used in the test.

\begin{tcolorbox}[colback=blue!5,colframe=blue!40!black]
\textbf{Finding 2:} PMT allows for coherence across results, making sure the outcome of the test will be similar for mutations that exhibit similar posterior distributions/similarity ratios contrary to MT.
\end{tcolorbox}

\textbf{Granularity:} Finally, note that using PMT, one can quantify if a mutation operator is more or less \textit{likely} to be killed or not killed. For instance, in Figure \ref{fig:post-f}, for UnityEyes (UE) $TCL - 3.12$, the posterior distribution is similar to the one of the original (\ie healthy) model and exhibits a similarity ratio of \textbf{0.74}, thus it surely would not be killed by the test set no matter the instances and so there is a very strong proof for considering the mutation \textit{likely} not killed. This analysis is not something MT would tell us, as it returns a deterministic decision over the given instances. Similarly, for $TCL - 6.25$, we only have a weak effect to consider the mutation \textit{likely} not killed, which is still more than for $TCL - 9.38$ where the effect is negligible with a ratio of \textbf{1.00}. As such, if there are some incentives to say that indeed $TCL - 6.25$ can be considered \textit{likely} not killed (and thus there is evidence that it should not be considered killed), such incentives do not exist for $TCL - 9.38$. Nonetheless, MT would just consider those two mutations to be similarly \enquote{not killed}, which limits the potential to analyze them. Similarly between $TCL - 12.5$ and $TCL - 18.75$, which are both considered killed by MT, yet our approach highlights that we have more incentives to consider the mutation \textit{likely} killed for the latter rather than for the former.

\begin{tcolorbox}[colback=blue!5,colframe=blue!40!black]
\textbf{Finding 3:} PMT delivers a finer grain analysis of the mutations, which allows to compare them on the likeliness of the mutation being killed. This might enable, for instance, to adjust potential thresholds one user would select to consider a mutation \textit{likely} killed, depending on the similarity ratio obtained. This is not possible with MT, which will consider all the killed mutations as similar.
\end{tcolorbox}

\subsubsection{PMT trade-off study}\label{sec:results-trade-off}

\begin{figure*}[]
\centering
\begin{minipage}{.45\linewidth}
    \centering
    \includegraphics[width=\linewidth,height=0.85\textwidth]{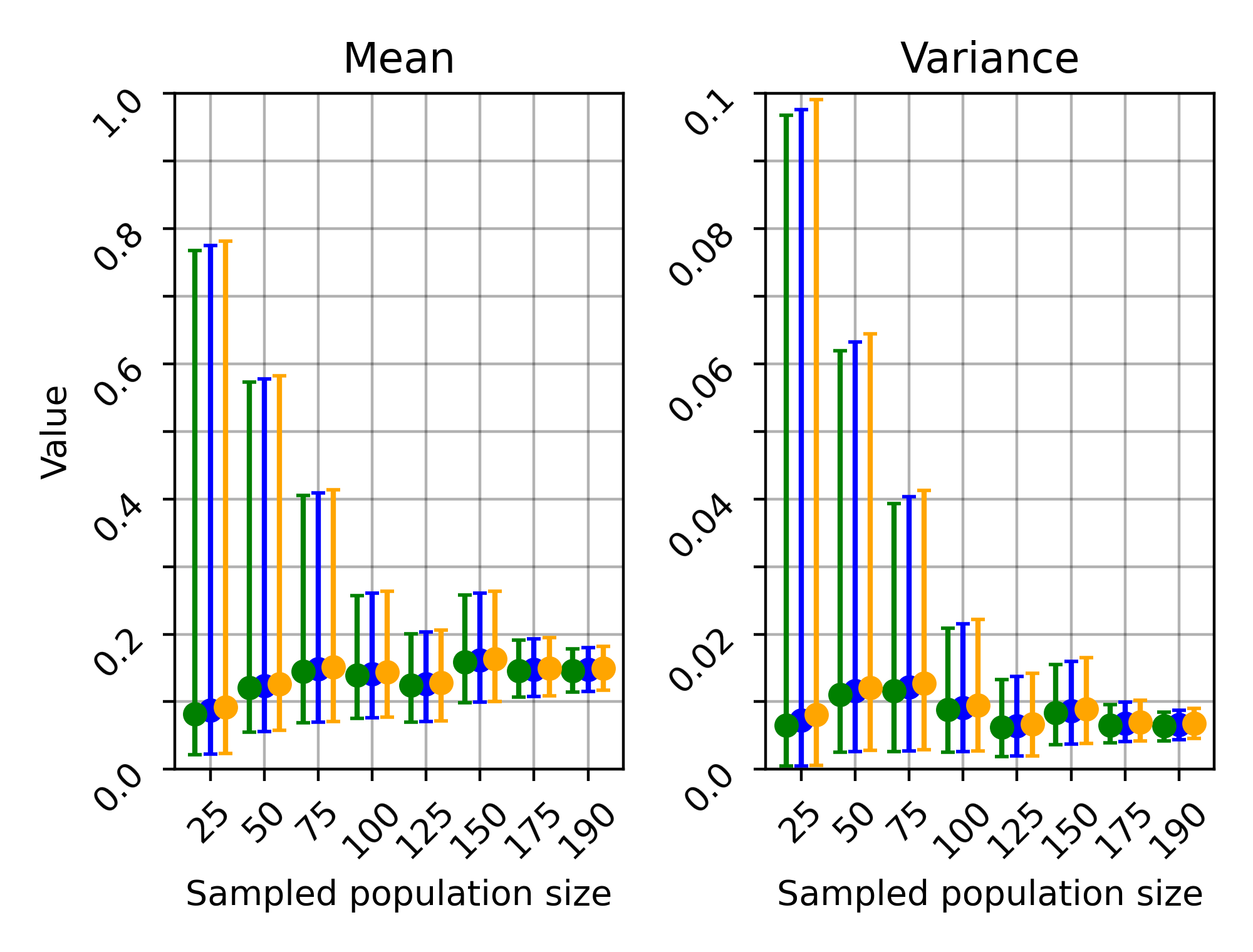}
    \includegraphics[width=\linewidth,height=0.85\textwidth]{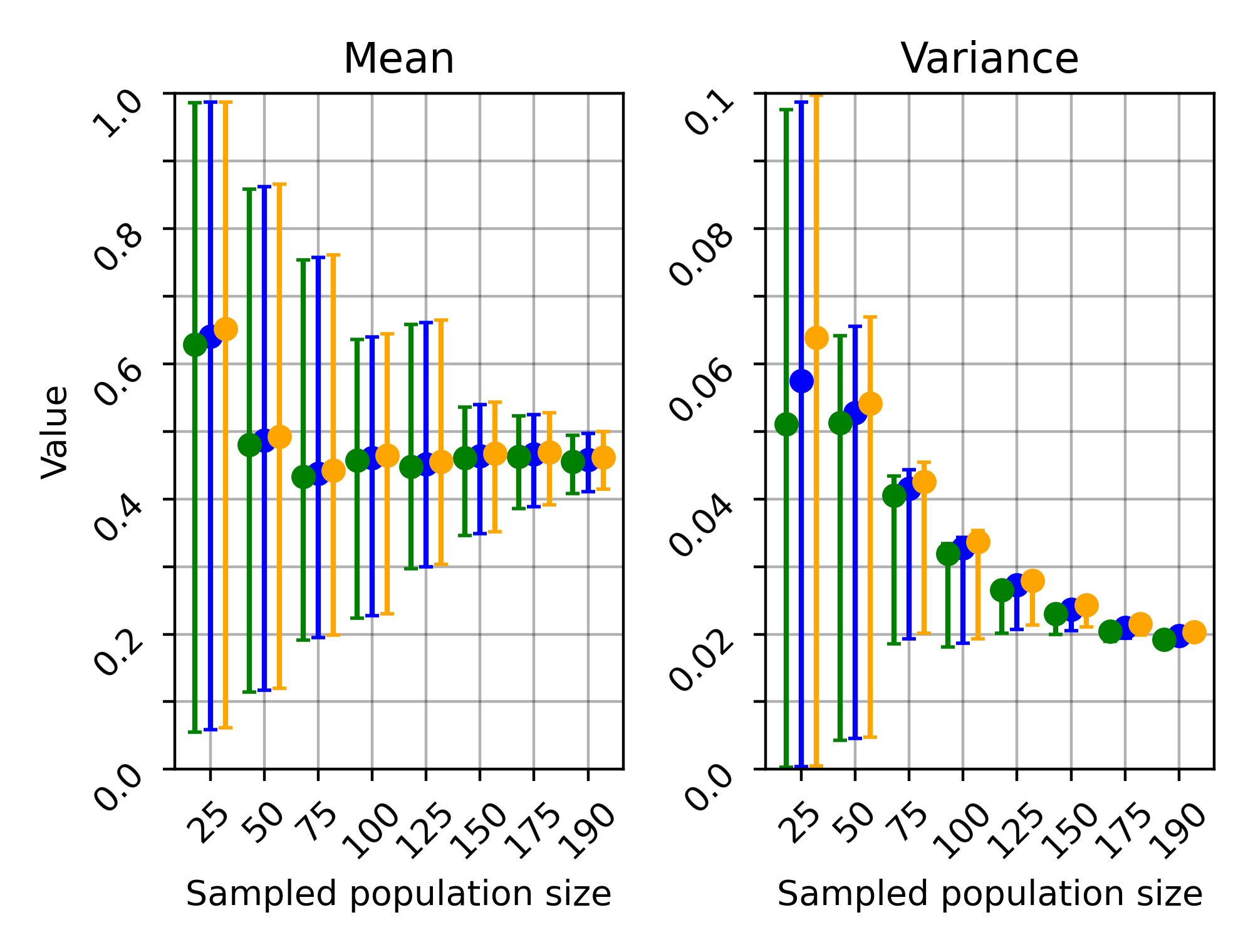}
    \includegraphics[width=\linewidth,height=0.85\textwidth]{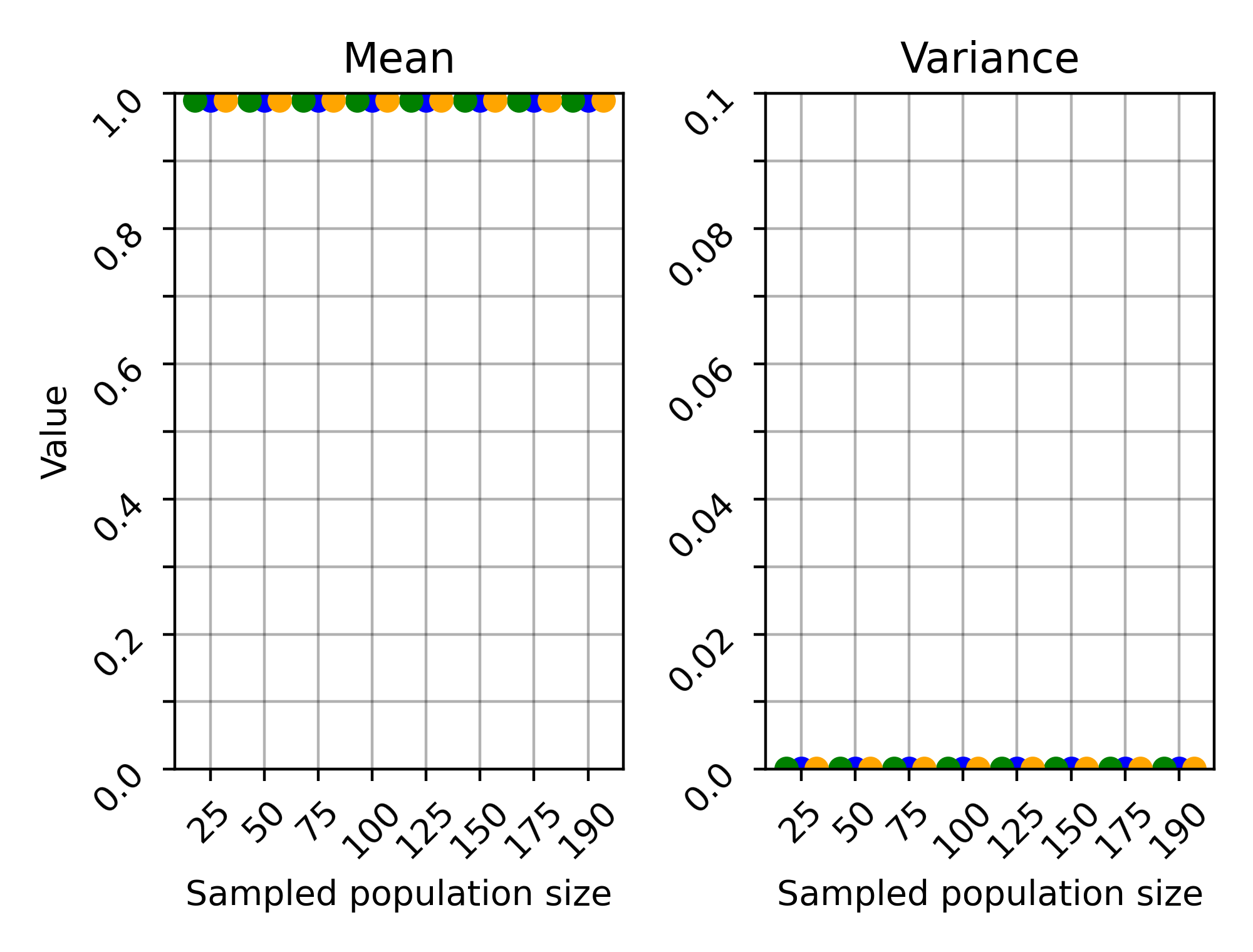}
  \end{minipage}\quad
  \begin{minipage}{.45\linewidth}
    \centering
    \includegraphics[width=\linewidth,height=0.85\textwidth]{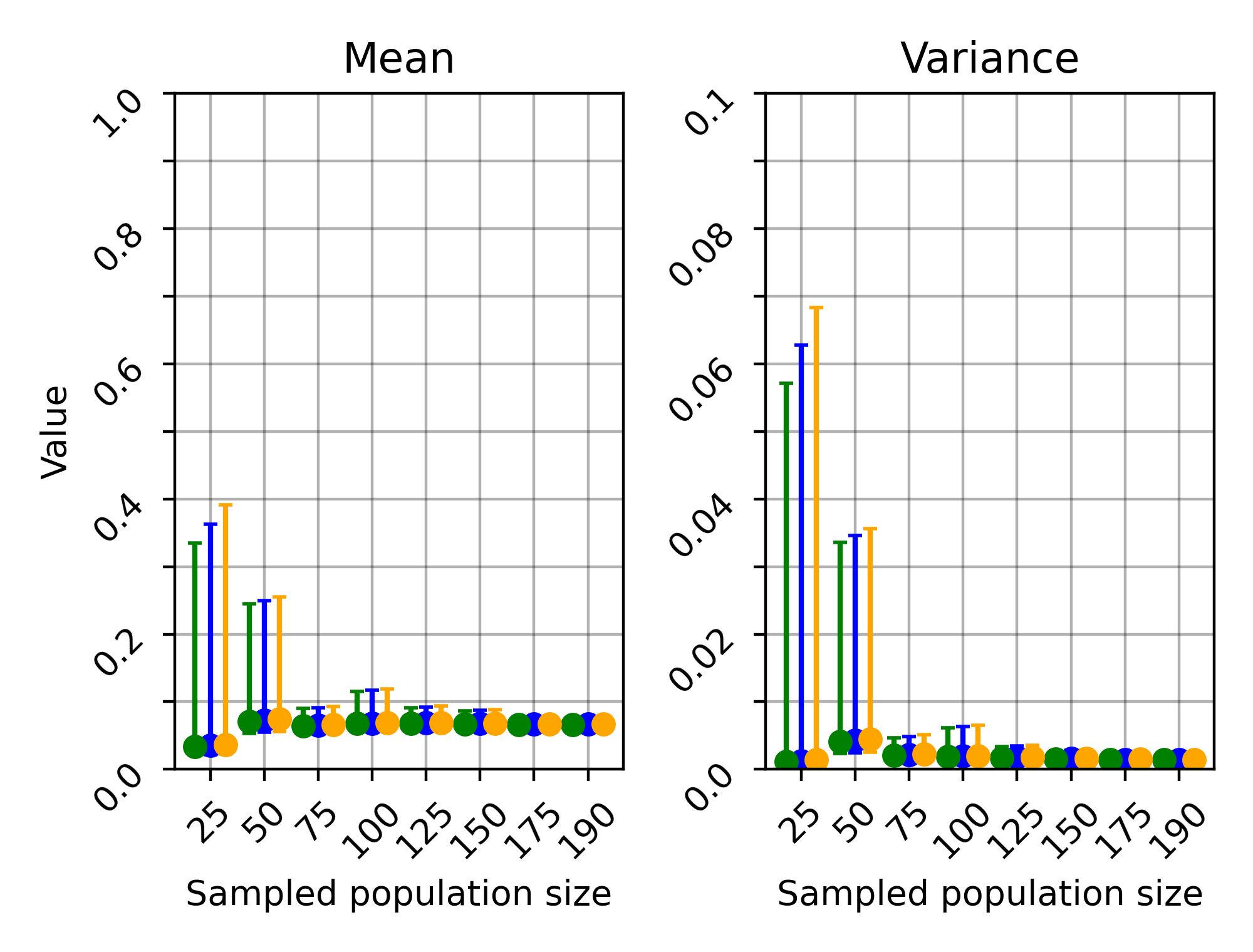}
    \includegraphics[width=\linewidth,height=0.85\textwidth]{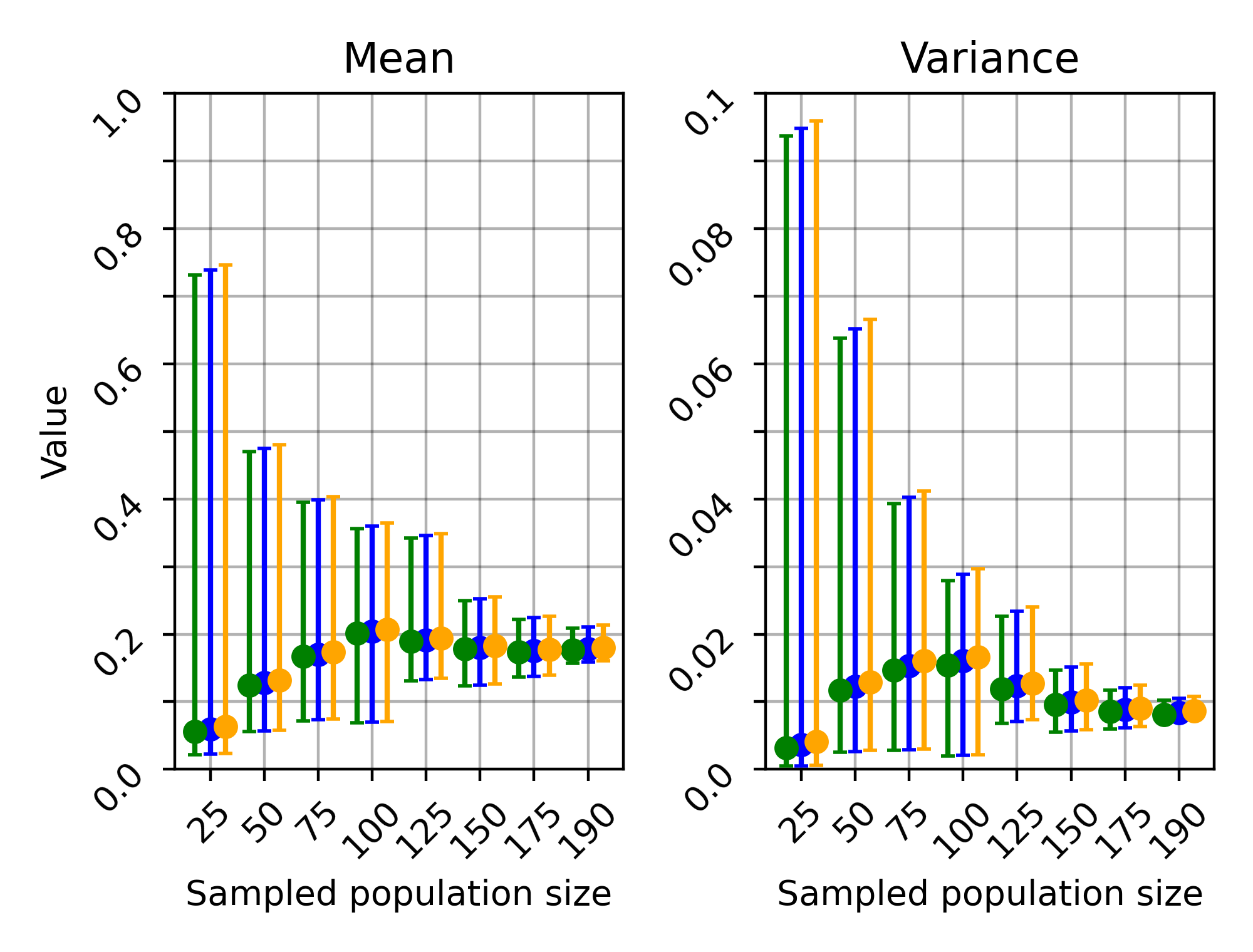}
    \includegraphics[width=\linewidth,height=0.85\textwidth]{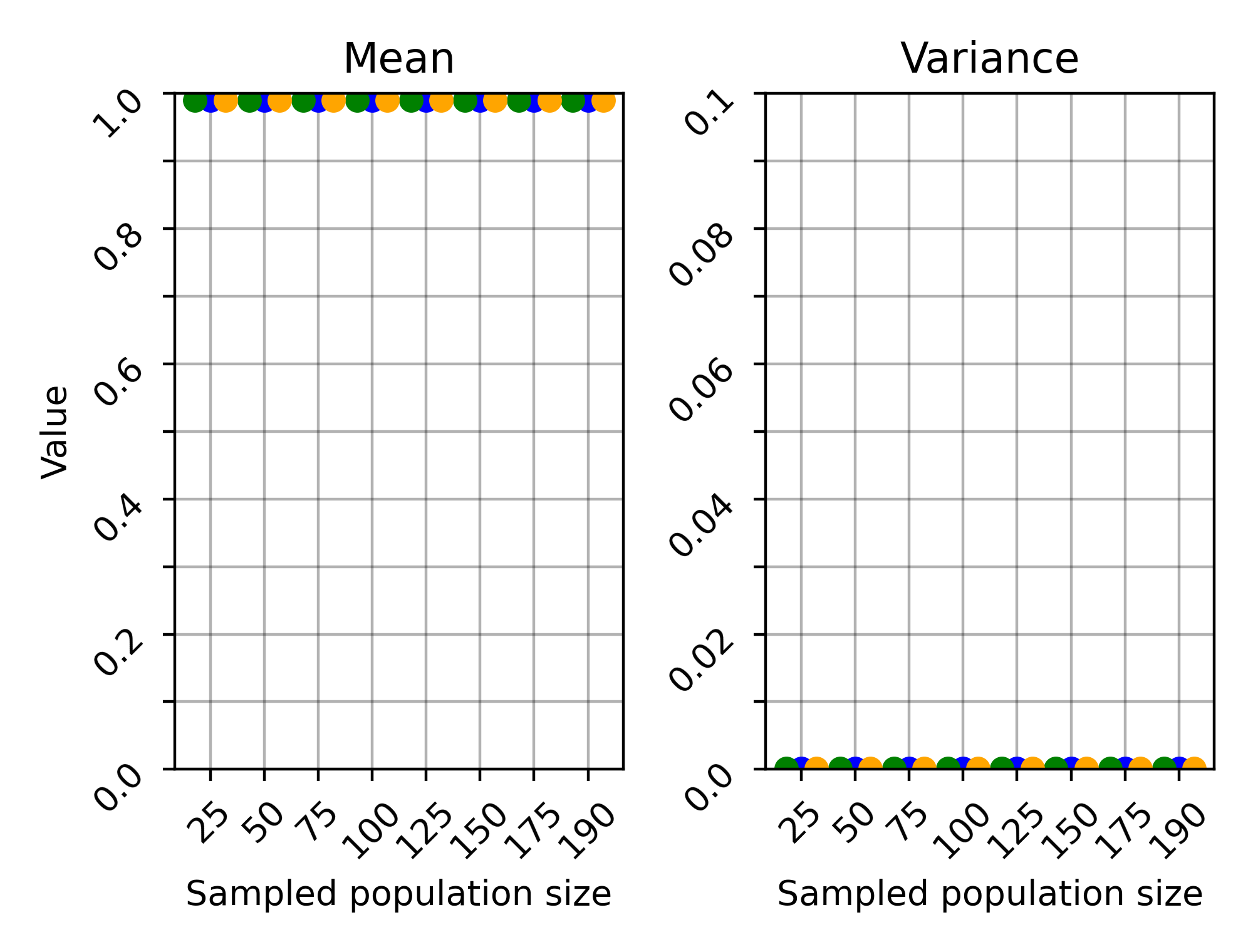}

    \end{minipage}
\caption{Error of estimates when testing on \enquote{delete\_training\_data} (left) and \enquote{change\_weights\_initialisation} (right) mutation operators for MNIST on three magnitudes (top to bottom: $3.1$, $9.29$, $30.93$ and \textit{glorot normal}, \textit{he normal} and \textit{zeros}). For each estimate, we present the average over $N_{pop} = 30$ of the monte-carlo estimate (\textcolor{blue}{blue}), monte-carlo lower bound (\textcolor{green_ok}{green}) and monte-carlo upper bound (\textcolor{orange}{orange}) as calculated in Section \ref{sec:methodology-error}. We also display the 95\% confidence interval.}
\label{fig:trade_off_mnist}
\end{figure*}

\begin{figure*}[]
\centering
\begin{minipage}{.45\linewidth}
    \centering
    \includegraphics[width=\linewidth,height=0.85\textwidth]{images/mnist_delete_training_data_3.1_std.png}
    \includegraphics[width=\linewidth,height=0.85\textwidth]{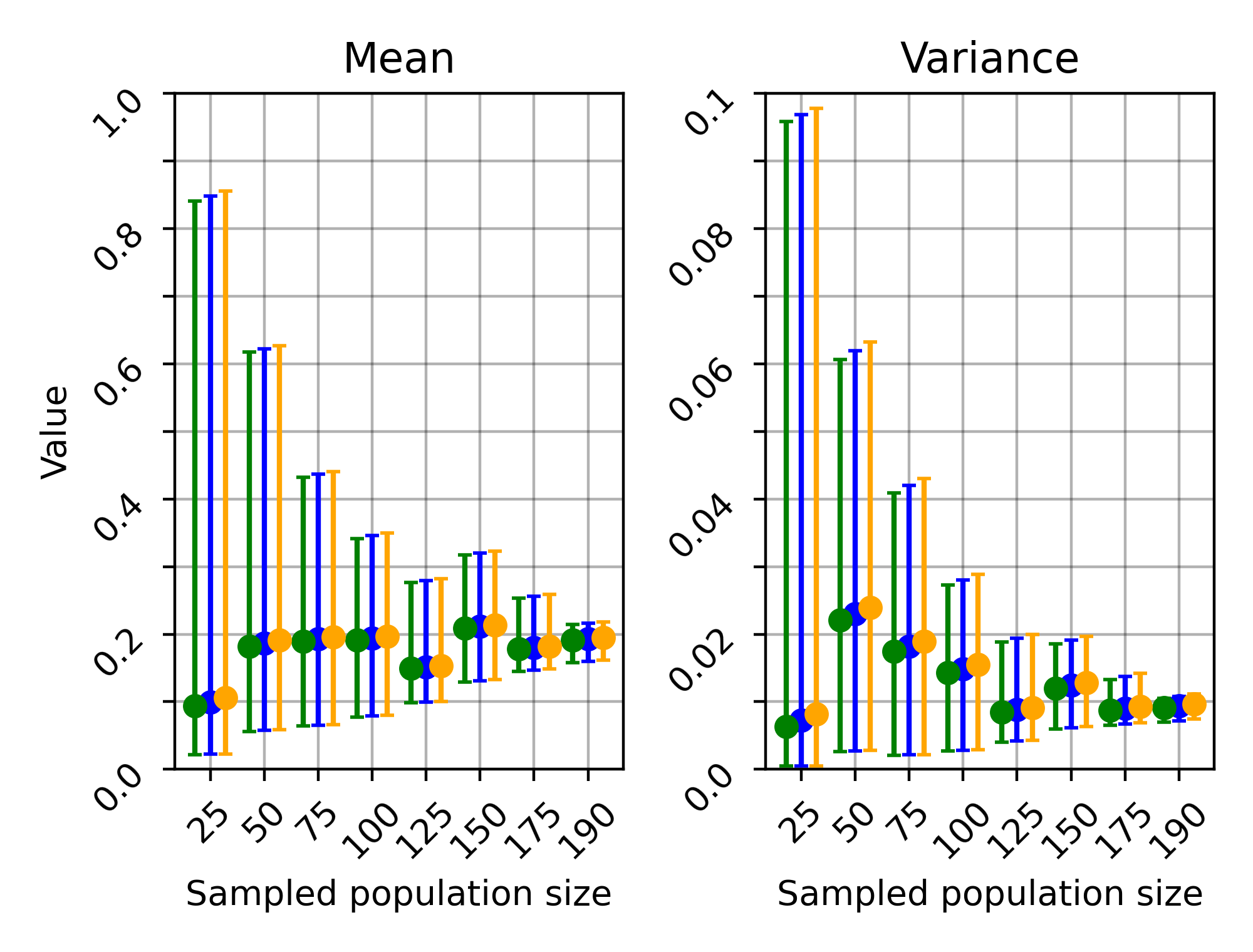}
    \includegraphics[width=\linewidth,height=0.85\textwidth]{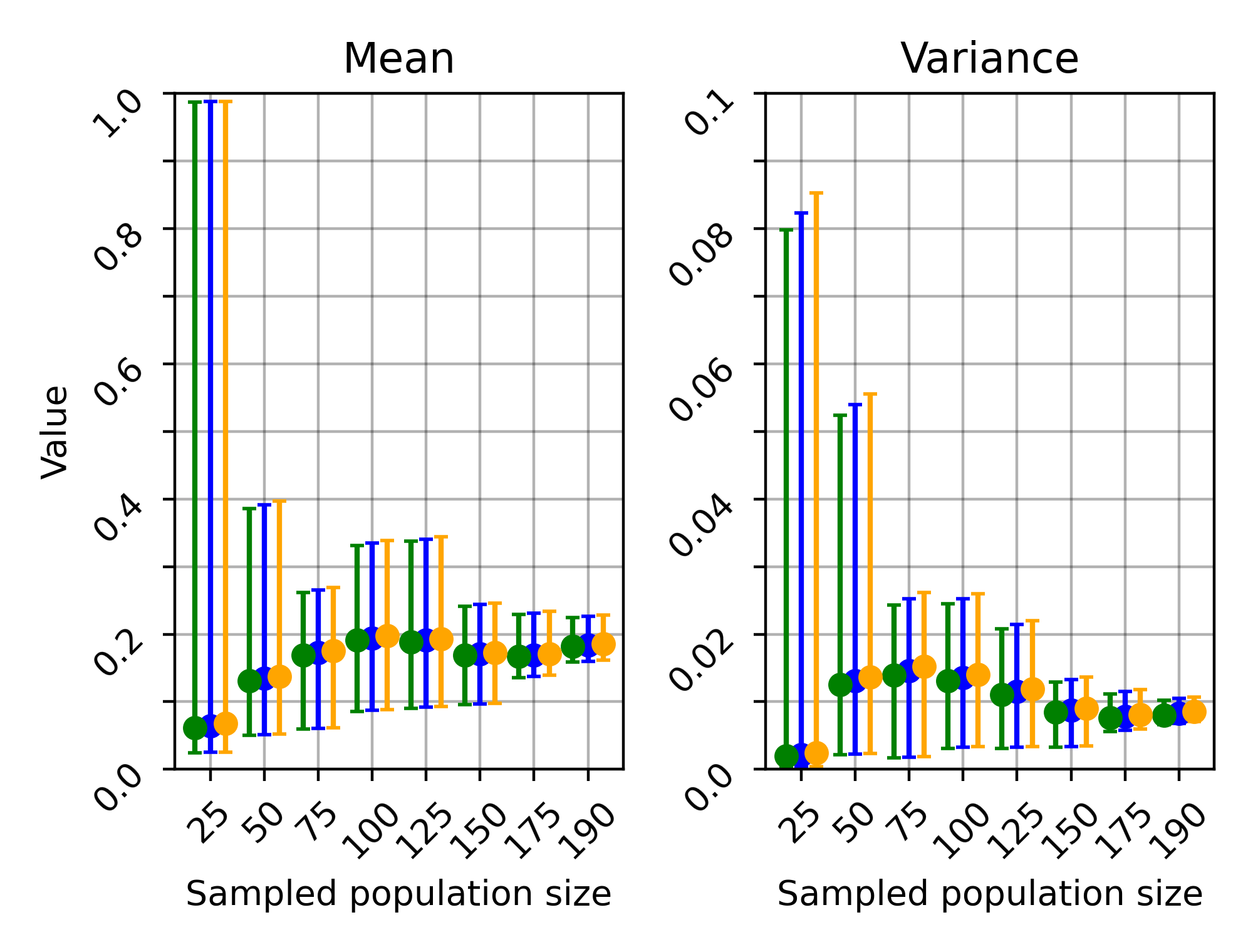}
  \end{minipage}\quad
  \begin{minipage}{.45\linewidth}
    \centering
    \includegraphics[width=\linewidth,height=0.85\textwidth]{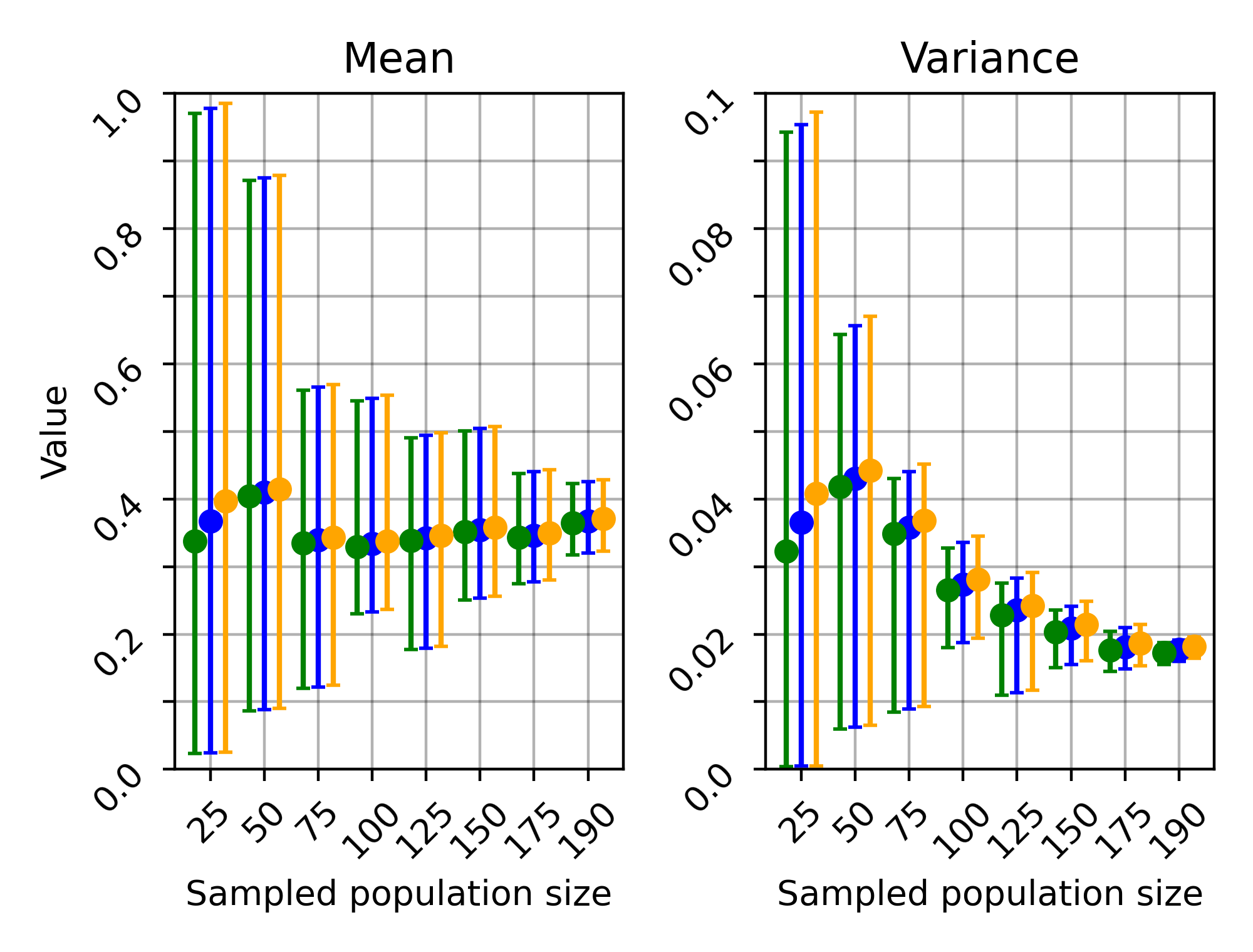}
    \includegraphics[width=\linewidth,height=0.85\textwidth]{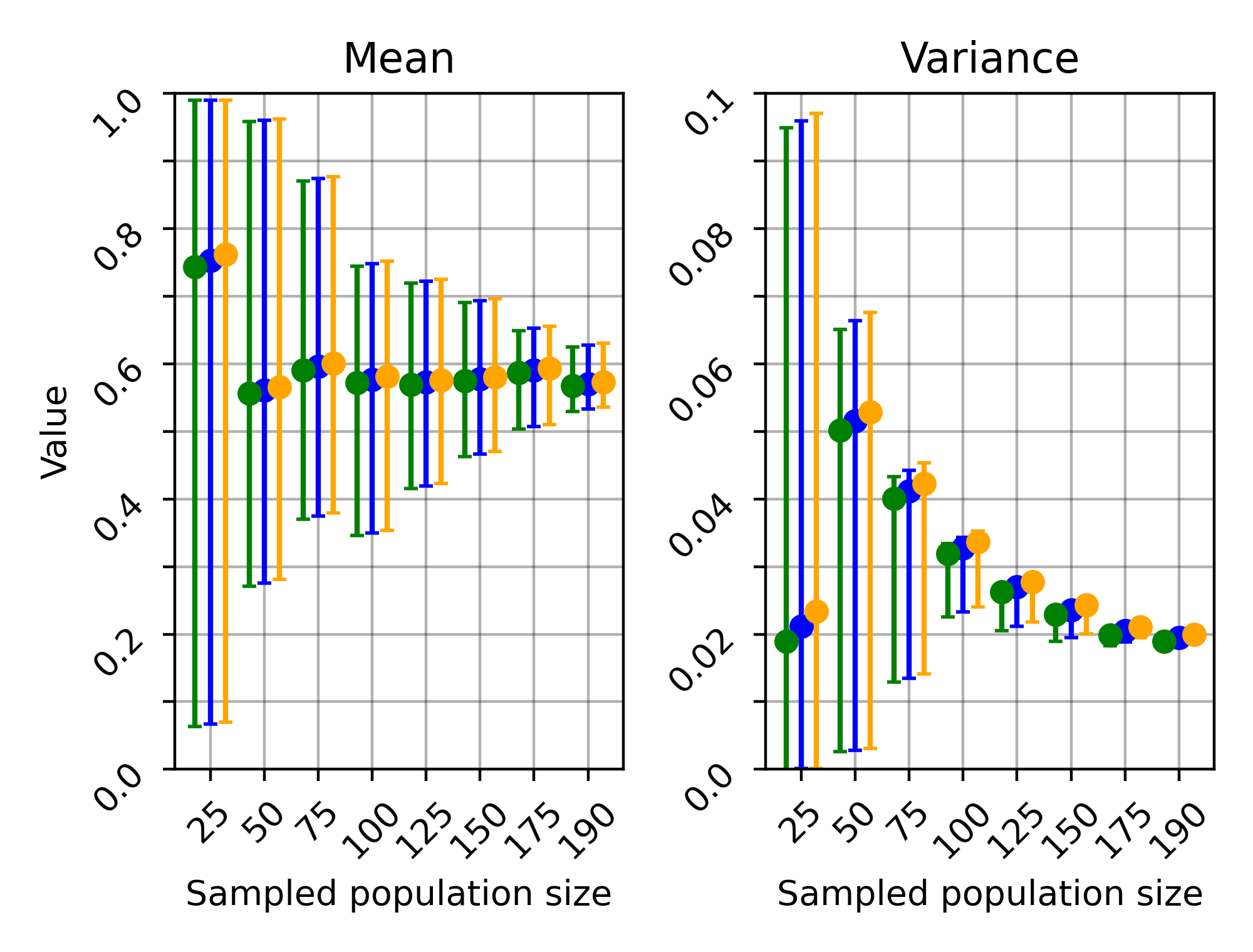}
    \includegraphics[width=\linewidth,height=0.85\textwidth]{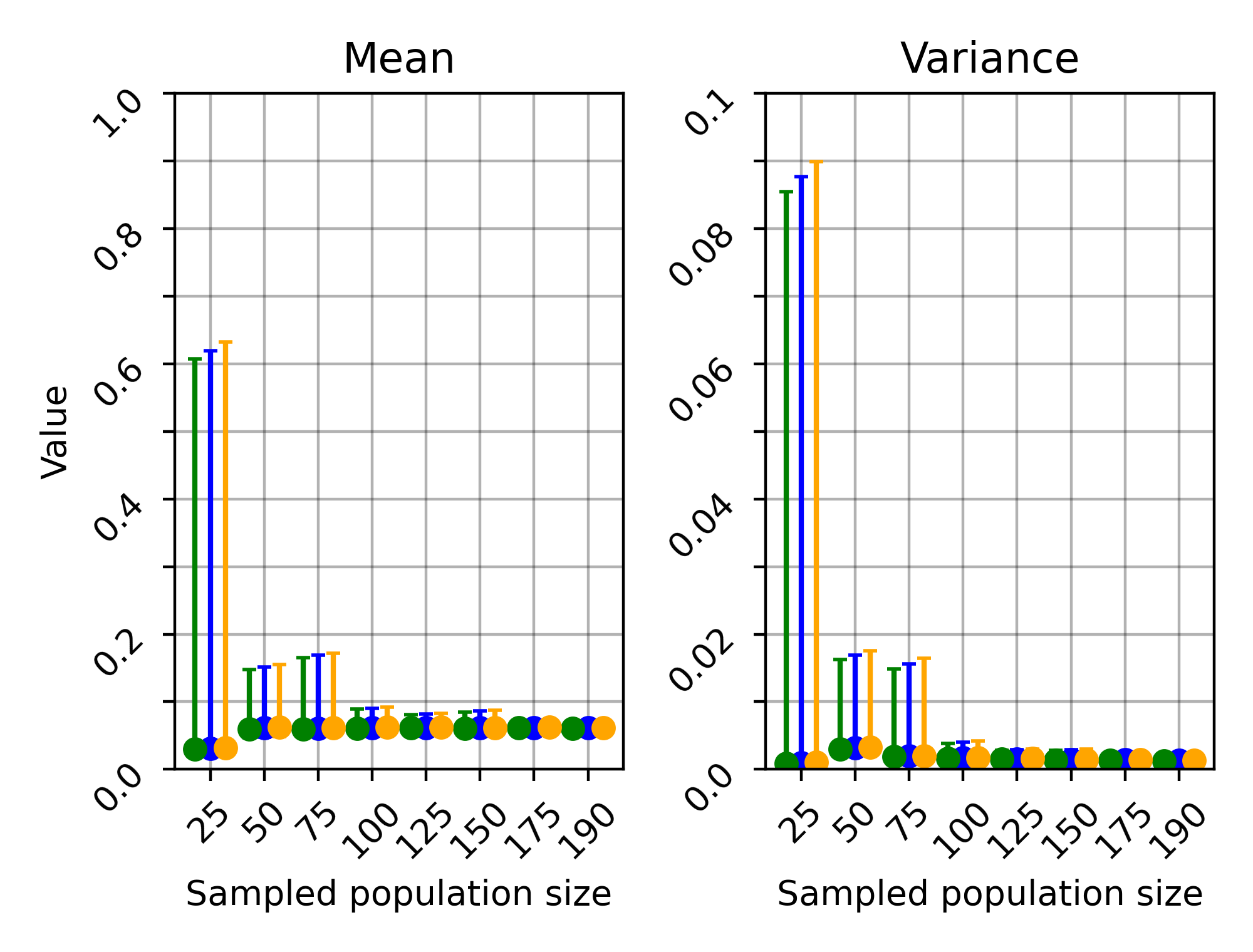}

    \end{minipage}
\caption{Error of estimates when testing on \enquote{delete\_training\_data} (left) and \enquote{change\_label} (right) mutation operators for the three models the same magnitude (top to bottom: $MNIST$, $MovieRecomm$, $UnityEyes$). For each estimate, we present the average over $N_{pop} = 30$ of the monte-carlo estimate (\textcolor{blue}{blue}), monte-carlo lower bound (\textcolor{green_ok}{green}) and monte-carlo upper bound (\textcolor{orange}{orange}) as calculated in Section \ref{sec:methodology-error}. We also display the 95\% confidence interval.}
\label{fig:trade_off_mnist2}
\end{figure*}

Figure \ref{fig:trade_off_mnist} and \ref{fig:trade_off_mnist2} show the error estimation of the estimates when computing the bagged posterior with different samples and different sample sizes. Figure \ref{fig:trade_off_mnist} focuses on one model/mutation operator and varies the magnitude of the mutation, while Figure \ref{fig:trade_off_mnist2} shows the results for the same mutation operator and same magnitude for different models. Similar trends can be noticed for other models/mutation operators. First, from these graphs, we can make the general following observations:
\begin{itemize}
    \item The larger the sample size, the lower the error across the different samples of the same size. This resonates with the intuition that the bootstrap hypothesis is increasingly valid. In other words, the sample is increasingly more representative of the unknown underlying population as we increase the number of instances in the sample.
    \item The average across the $N_{pop} = 30$ runs of the different estimates, as well as the average of their lower bound and upper bound (dot on the plots), are close. This suggests that for a given sample, the individual confidence interval, on average, is not large. In a nutshell, this means that there is not a huge variation between the bagged posterior obtained from the monte-carlo simulation for a given sample. Overall, our findings suggest that there is a low monte-carlo error when estimating the bagged posterior approximation error for $B = 100$ bootstrapped datasets, similarly to Huggins \cite{Huggins19} observations.
\end{itemize}  

As a consequence of these findings, if the bagged posterior approximation error is low, the error due to the representativity (and so the size) of the sampled population is big, yet it will decrease logically as the sample size increases. Of course, the larger the number of available trained instances the better. In practice, it seems that our choice of $200$ instances is indeed warranted, as the variation across samples decreases with the sample size, and for $190$ the confidence intervals are relatively small.

\begin{tcolorbox}[colback=blue!5,colframe=blue!40!black]
\textbf{Finding 4:} When applying PMT, following Huggins \cite{Huggins19} observation of setting $B = 100$ for the bootstrapped repetitions is a sound choice. Moreover, our choice of leveraging a sampled population of size $200$ also seems warranted as the approximation error over the bagged posterior is relatively small with a sample size of $190$.
\end{tcolorbox}

Secondly, we can now compare the evolution of the error estimation across models and mutations. In Figure \ref{fig:trade_off_mnist} we can compare the evolution through the increased magnitude of the mutations. We note for instance that the error estimation tends to be lower for a mutation operator with a low or high magnitude compared to a medium one ($3.1$ and $30.93$ vs $9.29$, \textit{glorot normal} and \textit{zeros} vs \textit{he normal}). Most likely, mutations with medium magnitude (for a given mutation operator) are more prone to divergence among the instances and more likely to have larger differences across samples. In Figure \ref{fig:trade_off_mnist2}, we compare the error estimation for the same mutation operator and magnitude across the models. There does not seem to be necessarily a similar evolution across models for the same mutation operator (see for instance \enquote{change\_label}, with \textit{UnityEyes} and the others), as such, the error estimation does not seem to be based on the mutation operator, but rather to be model dependent.

\begin{tcolorbox}[colback=blue!5,colframe=blue!40!black]
\textbf{Finding 5:} Medium magnitude mutation operators tend to have a higher error for the same sampled size when compared to mutation operators with low/high magnitude. Moreover, there is no explicit trend in the decrease of the error for the same mutation operator across the different models, so the error reduction seems more model-dependent.
\end{tcolorbox}

\subsection{Discussion}

In Section \ref{sec:results}, we have shown that using a deterministic test over a set of instances will not offer stable results for MT, hence the flakiness issue we mentioned. On the contrary, using PMT to calculate the posterior distribution of the test allows for better insights into the mutation operators under test. One important point is that, as mutations can be both killed and not killed by a test for some particular instances, the notion of killing a mutation as used in traditional MT does not seem to be very relevant in the context of ML. As such, we prefer to refer to the notion of mutation being \textit{likely} killed, \ie do we have sufficient evidence in a direction to assert it, similarly to how \textit{effect size} criterion would be used in a statistical test.

One consequence of this decision led us to introduce the similarity ratio metric $\mathcal{R}$ which allows for practical decision over the calculated posterior. We showed in the results of Section \ref{sec:results-exp1} that this metric yields a more insightful and finer grain analysis than the simple binary outcome of current MT frameworks. In particular, with PMT, we are able to get stable and coherent test results over the mutation operators, which would not be the case with MT because of the flakiness stemming from the selection/choice of the instances. We proposed an empirical scale to quantify the effect given by $\mathcal{R}$, based on scales that are used for \textit{effect size} in statistical test. This scale gives a rough idea to the user of the level of confidence attached to \textit{how likely} the mutation is killed, which can help the user make a decision about whether or not to consider the mutation \textit{likely} killed. In practice, a conservative choice would be to accept only mutations with a $\mathcal{R}$ of at least $1.15$, that is a \textit{strong} effect, which results in posterior distributions being very similar to the ideal mutant posterior. For instance, taking the mutation $MN - TRD$ we used in the motivating example of Section \ref{sec:motiv_ex}, using the results from Table \ref{tab:exp_1}, we would consider only the mutation of magnitude $30.93$ to be killed, as the effect is \textit{very strong}, the rest of the mutations being by default considered as not killed in order not to have false positive.

Note that, aside from a ratio $\mathcal{R}$ around $1$ or below (above) $0.82$ ($1.22$), intermediate levels of the scale might be regarded as arbitrary. Yet, they serve the practical purpose of allowing to at least be able to compare the effect of the different mutations, in a more meaningful way than the binary outcome of MT and in a more direct way than the more complicated analysis of the posterior distributions obtained through the bagging process.

%\Foutse{i think we should have a research question about this and move this content there!}\Florian{What do you mean? It's just a consequence of having to use more instances for the process, so it should be fine to just mention it, no?}
Although PMT requires more computations since it needs access to more training instances to obtain a stable posterior, it is fully automated, can be easily adapted to any new mutation/models using our provided framework inside the detailed replication package \cite{rep_pack} and, given trained instances, PMT does not have a huge time overhead ($\sim$ 1 minutes/mutations to make a decision), especially using parallelization. Our results have been computed with $200$ instances and we found the error to be relatively small for a sample size of $> 190$ when verifying empirically the MCE over the estimates. In practice, a lower number of instances may suffice depending on the precision required and the mutation operator/model under test.

In any case, we believe the ability to better analyze mutations in DNN settings (in particular, to avoid potential tests yielding that a mutation is killed when it's not) out-weights the increase in cost due to the higher number of instances needed. This is especially true for DNNs used in safety-critical systems, where the reliability of tests is crucial. %such as safety-critical ones.

\section{Threats to Validity}\label{sec:threats}

\textit{Construct validity.} PMT relies on some approximations in which error is empirically evaluated. As such, there is an intrinsic error that we cannot reduce to theory and which depends on the model/dataset/mutation used. The rest of the assumptions are grounded in theory or previous research works. However, we showed empirically that for a sufficient number of instances, the error is relatively small and thus does not impact much the decision.

Regarding the empirical scale, if it is mostly based on our experimental results, its main purpose is to allow us to draw a fair comparison among different mutations, in order to assess their different effects. Moreover, the scale was designed to be a direct way of interpreting and comparing the posterior distributions our approach was built, in a more practical way for the user. As such, the absolute value of the scale is less important than the relative comparison we can draw from it.

\textit{Internal validity.} Because of the computation overhead induced by our method, a high number of instances are needed for each mutation operator, we had to choose which mutation to evaluate, and on which model. As such, the choice of the mutations and models could have an impact on the results. To mitigate this threat, we made sure to choose mutation operators based on DeepCrime's Killability / Triviality analysis performed for their mutation operator. For Model level mutation, we used mutation operators listed in both DeepMutation and MuNN. Regarding mutation parameters, we used parameters provided in the replication package of DeepCrime \cite{Humbatova21} and the one mentioned in MuNN \cite{Shen18}. We also made sure to keep similar mutations across models/datasets, to allow for a point of comparison.

For the specific choice of models/datasets, we chose the models/datasets used in DeepCrime as they provided them in their replication package, along with mutation operators which were designed to work on such models, which is more practical and ensure better replication ability. The particular choice of models/dataset was then motivated by the number of epochs to reduce the computation overhead, but we made sure to choose diverse enough subjects (regression and classification, image-based and non-image-based...) to improve generalization.

Finally, note that although we have used the same mutation test as DeepCrime because it is the latest MT approach designed and because we leveraged their instances for comparison, the mutation function could be anything the user deems fit, as long as it respects the definition provided in Section \ref{sec:intro}. As such, PMT is general enough to be applicable in a wide variety of scenarii.

\textit{External validity.} We chose the same models/datasets as in DeepCrime, based on the popular framework Keras, which may limit the generalization of the study. Yet, it was necessary to ensure that we use mutations, models, and datasets in the same way as DeepCrime, to draw a fairer comparison as we used their instances, mutation test, and some of their operators. Nonetheless, we expect the results to generalize because the process is independent of the model/dataset/mutation used.

\textit{Reliability validity.} Not to overwhelm the paper with results and due to space limitations, we did not include all the results of our experiments. Therefore, we have provided the complete results in our replication package. We also provide all necessary details required to replicate our study, as well as the implementation of PMT in our replication package \cite{rep_pack}.

\section{Related works}\label{sec:related}

%\Foutse{we dont need to define DNN here! you already did in the problem formulation section!}DNN can be described as an ensemble of neuron layers, each neuron being a weighted function generally composed of a non-linear activation function, which allows tackling complex tasks. The specificity of DNN is that, compared to traditional programs, their logic is not coded but \enquote{learned} through a training process using data. Thus, there is inherent stochasticity to them, which is limiting if we are to employ techniques such as MT as we described in the introduction.

MT is an established technique in SE \cite{Schuler09, Baker12}. It has also been applied in settings where non-determinism is present, for instance, Probabilistic Finite Sates Machine \cite{Hierons07}. Recently, researchers have been applying MT to DNN, %As such it is not surprising that it would be applied to DNN,
with Nour et al.  \cite{Nour19} evaluating MT tool's effectiveness on DNN, and DeepMutation \cite{Ma18}, DeepMutation++ \cite{Hu19}, or MuNN \cite{Shen18} proposing MT framework specific for DNN. These approaches notably distinguished between \textit{source-level} mutations, \ie mutation acting on model \textit{before training} (for instance, removing part of the training data), and \textit{model-level} mutations, \ie mutations acting on an \textit{already trained} model (for instance, adding noise to weights of the model). However, DeepMutation and similar approaches do not necessarily take the stochastic nature of DNN into account and do not offer real faults-based mutation operators. Based on this observation,  Jahangirova et al. \cite{Jahangirova20} introduced the statistical mutation test we described in Section \ref{sec:intro} and compared empirically previous MT frameworks. This work was then further extended leading to DeepCrime \cite{Humbatova21}, where authors proposed a training set-based analysis of killability and redefined the notion of triviality using fuzzy logic. Although they proposed new measures for mutation test analysis, they still used the same statistical test as in the previous work, which resulted in the flakiness we described in Section \ref{sec:intro}. In this paper, we propose a novel formulation of %ew way to describe
MT for DNN. Our proposed approach allows for stable test results, mitigating the flakiness issue, as well as a finer grain analysis of the mutations' behavior which makes possible a more coherent decision over the test results across mutations.

Note that there is still a debate on %MT application in DNN, notably
what constitutes an acceptable mutation in DNN. % is still up to debate.
For instance, according to Panichella et al. \cite{Panichella21}, source-level mutations such as those evaluated in DeepCrime may not be regarded as mutations in the classical sense. The argument is that %, they consider that,
since DNN can be seen as a test-driven development procedure and the training data as a test suite, source-level mutation operators (for instance, removing a percentage of train data) affect the test suite rather than the production code. This interpretation is up to debate since the training data is a crucial part of a DNN specification \cite{Gauerhof20, Salay18} and not simply a part of a test suite. In fact, the training data is responsible for what the DNN \enquote{learns} contrary to a simple test data that evaluates what the DNN has learned. Hence, mutating training data can be considered similar %is equal
to mutating the production code, since modifying the specification leads to a different model.
In this paper, similarly to \cite{Humbatova21, Jahangirova20}, we consider mutations over the training process to be proper for MT.

\section{Conclusion}\label{sec:conclusion}

This paper introduced PMT, a probability framework for MT in deep learning, to solve the flakiness inherent to current MT approaches. Using real-faults-based mutations as well as the mutation test proposed by DeepCrime, we evaluated PMT, showing how to leverage it to decide whether a mutation can be considered killed or not, in a more reliable way than what was proposed before.  Moreover, we showed that, for a sufficient number of instances, the approximation made in PMT (approximate bagged posterior and sample size effect) can be neglected, effectively stabilizing the posterior obtained by the process. Finally, the approach is fully automated, the decision does not introduce a huge time overhead once instances are trained and can be extended easily to any model/dataset/mutation.

In future work, we plan on investigating how to reduce the higher number of instances needed by PMT, possibly by evaluating if it is possible to predict the mutation behavior.

\section*{Funding Sources}

This work is supported by the DEEL project CRDPJ 537462-18 funded by the National Science and Engineering Research Council of Canada (NSERC) and the Consortium for Research and Innovation in Aerospace in Québec (CRIAQ), together with its industrial partners Thales Canada inc, Bell Textron Canada Limited, CAE inc and Bombardier inc.

%\clearpage

%% The Appendices part is started with the command \appendix;
%% appendix sections are then done as normal sections
%% \appendix

%% \section{}
%% \label{}

%% If you have bibdatabase file and want bibtex to generate the
%% bibitems, please use
%%
%%  \bibliographystyle{elsarticle-harv}
%%  \bibliography{<your bibdatabase>}

\bibliographystyle{elsarticle-num}
\bibliography{sample-base}
\end{document}